\documentclass[11pt]{article}
\usepackage{jheppub}

\usepackage{epsfig,multicol}
\usepackage{mathrsfs}
\usepackage{amsmath,amssymb,amsbsy,mathtools}
\usepackage{graphicx,picinpar,epsfig}
\usepackage{tikz}\usetikzlibrary{decorations.markings, arrows.meta}
\usepackage{calc}

\newcommand\fverb{\setbox\fverbbox=\hbox\bgroup\verb}
\newcommand\fverbdo{\egroup\medskip\noindent%
			\fbox{\unhbox\fverbbox}\ }
\newcommand\fverbit{\egroup\item[\fbox{\unhbox\fverbbox}]}
\newbox\fverbbox

\newcommand{\pslash}{p\kern-1ex /}
\newcommand{\qslash}{q\kern-1ex /}
\newcommand{\lslash}{l\kern-1ex /}
\newcommand{\sslash}{s\kern-1ex /}
\newcommand{\kaslash}{k_a\kern-2ex /}
\newcommand{\kbslash}{k_b\kern-2ex /}
\newcommand{\Dslash}{\mathcal{D}\kern-1.5ex /}

\newcommand{\beqa}{\begin{eqnarray}}
\newcommand{\eeqa}{\end{eqnarray}}

\newcommand{\ba}{\begin{eqnarray}}
\newcommand{\ea}{\end{eqnarray}}
\newcommand{\be}{\begin{equation}}

\numberwithin{equation}{section}

\title{Analytic resurgence in the O(4) model}

\author{
Zolt\'an Bajnok,\footnote{\tt bajnok.zoltan@wigner.hu}\ \  
J\'anos Balog\footnote{\tt balog.janos@wigner.hu}\ \ and \
Istv\'an Vona\footnote{\tt vona.istvan@wigner.hu}
\\
\vskip 1ex
{\it Holographic QFT Group, Institute for Particle and Nuclear Physics,\\
Wigner Research Centre for Physics} \\
{\it H-1525 Budapest 114, P.O.B. 49, Hungary}\\
}

\abstract{
We study the perturbative expansion of the ground state energy in the presence
of an external field coupled to a conserved charge in the integrable
two-dimensional O$(4)$ nonlinear sigma model.
By solving Volin's algebraic equations for the perturbative coefficients we
study the large order asymptotic behaviour of the perturbative series
analytically. We confirm the previously numerically found leading behaviour
and study the nearest singularities of the Borel transformed series and the
associated alien derivatives. We find a \lq\lq resurgence'' behaviour: the
leading alien derivatives can be expressed in terms of the original perturbative
series. A simplified 'toy' model is also considered: here the perturbative
series can be found in a closed form and the resurgence properties are very
similar to that found in the real problem.
}

\begin{document}


\newcommand{\con}{\,\star\hspace{-3.7mm}\bigcirc\,}
\newcommand{\convu}{\,\star\hspace{-3.1mm}\bigcirc\,}
\newcommand{\Eps}{\Epsilon}
\newcommand{\gM}{\mathcal{M}}
\newcommand{\dD}{\mathcal{D}}
\newcommand{\gG}{\mathcal{G}}
\newcommand{\pa}{\partial}
\newcommand{\eps}{\epsilon}
\newcommand{\La}{\Lambda}
\newcommand{\De}{\Delta}
\newcommand{\nonu}{\nonumber}
\newcommand{\beq}{\begin{eqnarray}}
\newcommand{\eeq}{\end{eqnarray}}
\newcommand{\ka}{\kappa}
\newcommand{\ee}{\end{equation}}
\newcommand{\an}{\ensuremath{\alpha_0}}
\newcommand{\bn}{\ensuremath{\beta_0}}
\newcommand{\dn}{\ensuremath{\delta_0}}
\newcommand{\al}{\alpha}
\newcommand{\bm}{\begin{multline}}
\newcommand{\fm}{\end{multline}}
\newcommand{\de}{\delta}
\newcommand{\dpd}{\int {\rm d}^d p}
\newcommand{\dqd}{\int {\rm d}^d q}
\newcommand{\dxd}{\int {\rm d}^d x}
\newcommand{\dyd}{\int {\rm d}^d y}
\newcommand{\dud}{\int {\rm d}^d u}
\newcommand{\dzd}{\int {\rm d}^d z}
\newcommand{\dpp}{\int \frac{{\rm d}^d p}{p^2}}
\newcommand{\dqq}{\int \frac{{\rm d}^d q}{q^2}}


\maketitle


\section{Introduction}

The two dimensional O$(N)$ $\sigma$ models provide an ideal playground
for theoretical particle and statistical physics, where non-perturbative
phenomena can be analyzed exactly without relying on approximate methods.
These models share many features with QCD including asymptotic freedom
in perturbation theory and the dynamical generation of a mass scale,
which breaks the classical scale invariance in the quantum theory.
These models were the first, where the exact S-matrix was determined
\cite{Zamolodchikov:1977nu}. By analysing the behaviour of the groundstate
energy density in a magnetic field \cite{Polyakov:1983tt} the perturbative
scale was the first time exactly related to the dynamically generated
mass \cite{Hasenfratz:1990zz,Hasenfratz:1990ab}. This energy density
then can be expanded in perturbation theory and its asymptotic series
revealed non-perturbative sectors \cite{Volin:2009wr,Marino:2019eym,Abbott:2020mba,Abbott:2020qnl}
related to renormalons, which exist also in QCD \cite{Bauer:2011ws}.
The theory which describes how the large order asymptotics coefficients
``resurge'' themselves into the non-perturbative contributions is
called resurgence\footnote{See
\cite{Marino:2012zq,Dorigoni:2014hea,Aniceto:2018bis} for introductions
and references.}, which aims at constructing the full non-perturbative ambiguity free
trans-series for the physical observables. The exact solvability of
the O$(N)$ $\sigma$ models gives a hope to establish the complete
resurgence structure exactly and to provide the sought trans-series,
which could indicate similar behaviours for analogous realistic theories,
such as for QCD. 

The resurgence program has been analyzed in many two dimensional and
related quantum field theories including also supersymmetric ones
\cite{Aniceto:2011nu,Aniceto:2014hoa,Aniceto:2013fka,Aniceto:2015rua,Arutyunov:2016etw,Demulder:2016mja,Dorigoni:2015dha,Dorigoni:2017smz}
and some of them focuses on the semiclassical domain of O$(N)$ and
principal chiral models \cite{Dunne:2015eaa,Dunne:2015ywa,Dunne:2016nmc,Fujimori:2018kqp}.
Recently there have been active research and significant progress
on relativistic integrable quantum field theories \cite{Marino:2019eym,Marino:2021six,DiPietro:2021yxb}
and their non-relativistic statistical physical counterparts \cite{Marino:2019fuy,Marino:2020ggm,Marino:2020dgc}.
Resurgence proved to be a useful tool also in non-integrable quantum
field theories, such as in the $\phi^{4}$ theory \cite{Serone:2018gjo,Serone:2019szm}
and also in analysing the hydrodynamics of the Yang Mills plasma \cite{Aniceto:2015mto,Aniceto:2018uik}.
There is no asymptotically free QFT, however, where the full resurgence
structure was derived. In this paper we would like to make the first
step into this direction by analysing the O$(4)$ model, which is
the simplest amongst the O$(N)$ theories. 

In the integrable O$(N)$ $\sigma$ models we can couple one of the
O$(N)$ conserved charges to a magnetic field and analyse the groundstate
energy density in two different ways \cite{Hasenfratz:1990zz,Hasenfratz:1990ab}.
On the one hand it can be calculated perturbatively in terms of $\Lambda_{\overline{MS}}$
for large magnetic fields. On the other hand, integrability enables
to write an exact integral equation for the energy density, which
can be expanded using Wiener-Hopf technics. The comparison of the
first order terms provided the celebrated mass/coupling relation.
Further perturbative terms turned to be decisive in the AdS/CFT correspondence,
where the O$(6)$ model plays a crucial role \cite{Bajnok:2008it}.
The calculation of higher order terms are cumbersome and the breakthrough
came from the integrable side, where Volin could manage to translate
the integral equation into a system of algebraic ones
\cite{Volin:2009wr,Volin:2010cq}.
These recursive equations could be solved for the first 20-50 terms
analytically, which allowed to observe a factorial growth \cite{Volin:2009wr,Volin:2010cq,Marino:2019eym}.
It also helped to identify the closest singularities on the Borel
plane, which corresponds typically to UV and IR renormalons, diagrams
in perturbation theory, where the asymptotic growths arise from integrating
over UV or IR domains in specific renormalon diagrams \cite{Beneke:1998ui}.
In the large $N$ limit the leading of such diagrams were identified
and their contributions were calculated exactly, which agreed with
the solution of the algebraic equations \cite{Marino:2021six}. 

In our recent works \cite{Abbott:2020mba,Abbott:2020qnl}, we focused
on the O$(4)$ model, where the algebraic equations simplified drastically.
We solved these equations numerically with very high precision, from
which we conjectured analytical expressions for their leading and
subleading factorial behaviours. We numerically identified logarithmic
cuts on the Borel plane. We investigated the functions multiplying
these cuts (alien derivatives) perturbatively from the asymptotics
of the original perturbative series. By using high precision numerics
we could identify and characterise further cuts on the real line
and with their help we could manage to write the first few terms
of an ambiguity free non-perturbative trans-series, which we confronted
with the numerical solution of the original integral equation and
we found complete agreement. The full power of the resurgence theory
can be used, however, when we can bridge the various non-perturbative
expansions to each other, as then we would have relations between
low order perturbative coefficients. Based on our numerics we conjectured
such relations for the nearest cuts on the Borel plane, i.e. for the
leading non-perturbative corrections. The aim of the present paper
is to elevate these numerical observations to an analytically proved
level and to prepare the ground for systematic analytical studies, which
could reveal the full resurgence structure. 

The paper is organized as follows: in the rest of the introduction
we collect the needed formulae from the literature. In particular,
we introduce the exact linear integral equation for the ground-state
energy density and Volin's approach for the corresponding resolvent,
which turns the integral equations into algebraic ones for the systematic
perturbative expansion. We simplify these recursive equations in the
O$(4)$ case and emphasize that the only input comes from the leading
behaviour of the resolvent's Laplace transformed, which can be determined
from the Wiener-Hopf solution. In the next section we summarize the
main results of this paper. Technically, we manage to formulate recursive equations
directly for the parametrizations of the Laplace transform of the
resolvent without relying on its coordinate space form. We then calculate
the asymptotic form of the perturbative expansions, which provides
the analytical solution for the resolvent's Laplace transform around
the origin. We then learn how to extend the asymptotic form selfconsistently,
which is valid on the unit circle and gives the complete analytical
information on leading singularities of the Borel plane, i.e. the
leading alien derivates. We summarize the main findings in the conclusion
where we also present an outlook for further research. All the technical
details are relegated to appendices.

\subsection{TBA calculation of the ground-state energy density}

The central object of study in this paper is the ground-state energy in a magnetic field $h$
\cite{Polyakov:1983tt,Hasenfratz:1990zz,Hasenfratz:1990ab}, when
the Hamiltonian is modified as $H(h) = H(0)-hQ_{12}$, where $H(0)$ is the original sigma model
Hamiltonian and $Q_{12}$ is the conserved charge associated to the (internal) rotation symmetry
in the $12$ plane. Particles with largest $Q_{12}$ charge condense into the vacuum if $h$ is larger
than $m$, the mass of the particles. Due to the integrability of the model this condensate can
be described by the Thermodynamic Bethe Ansatz (TBA) equation \cite{Polyakov:1983tt,Hasenfratz:1990zz,Hasenfratz:1990ab}
\begin{equation}
\chi(\theta)-\int_{-B}^{B}\frac{d\theta'}{2\pi}K(\theta-\theta')\chi(\theta')=\cosh\theta\label{TBA}
\end{equation}
where the kernel is determined by the scattering matrix $S(\theta)$ for the condensed particles:
\begin{equation}
2\pi K(\theta)=-2\pi i\partial_{\theta}\log S(\theta)=\sum_{k=\{\frac{1}{2},\Delta\}}\{\Psi(k+\frac{1}{2}-\frac{i\theta}{2\pi})
-\Psi(k+\frac{i\theta}{2\pi})\}+ {\rm cc}.
\end{equation}
Here $\Delta=\frac{1}{N-2}$ and $\Psi$ is the digamma function: $\Psi(\theta)=\partial_{\theta}\log\Gamma(\theta$).
The particle density $mU$ and the energy density $m^2W$ are obtained simply from 
\begin{equation}
U=\int_{-B}^{B}\frac{d\theta}{2\pi}\chi(\theta)\quad;\qquad W=\int_{-B}^{B}\frac{d\theta}{2\pi}\cosh\theta\,\chi(\theta).\label{rhoeps}
\end{equation}
These equations depend on $B$, which can be related to the magnetic field $h$.
Large magnetic fields correspond to large densities and large $B$, and in the following we will study
the $1/B$ expansion of the solution and the densities.

\subsection{Perturbative expansion of the TBA equation, Volin's method \cite{Volin:2009wr,Volin:2010cq}}

The calculations can be simplified by introducing the resolvent
\begin{equation}
R(\theta)=\int_{-B}^{B}d\theta'\,\frac{\chi(\theta')}{\theta-\theta'}
\end{equation}
and its Laplace transform 
\begin{equation}
\hat{R}(s)=\int_{-i\infty+0}^{i\infty+0}\frac{dz}{2\pi i}e^{sz}R(B+z/2).
\end{equation}
The latter is related to the Fourier transform of $\chi(\theta)$: 
\begin{equation}
\hat{R}(s)=2e^{-2sB}\tilde{\chi}(2is)\quad;\qquad\tilde{\chi}(\omega)=\int_{-B}^{B}e^{-i\omega\theta}\chi(\theta)d\theta.\label{16star}
\end{equation}
Then, using the $\chi(\theta)=\chi(-\theta)$ symmetry 
\begin{equation}
W=\int_{-B}^{B}\cosh\theta\chi(\theta)\frac{d\theta}{2\pi}=\int_{-B}^{B}e^{\theta}\chi(\theta)\frac{d\theta}{2\pi}
=\frac{1}{2\pi}\tilde{\chi}(i)=\frac{e^{B}}{4\pi}\hat{R}(1/2).
\end{equation}

Although it is possible to proceed for generic O$(N)$ models, in the following we restrict
our attention to the O$(4)$ model where $\Delta=1/2$ since formulas
are much simpler there.

Volin made the following two seminal observations. First, both $R(\theta)$ and $\hat R(s)$ can be
expanded for large $B$ and the form of the two asymptotic expansions are determined by the
TBA equation and the analytic properties of its solution. In rapidity space
\begin{equation}
R(\theta)=2\tilde{A}\sqrt{B}\sum_{n,m=0}^{\infty}\frac{c_{n,m}}{B^{m-n}(\theta^{2}-B^{2})^{n+1/2}},
\end{equation}
where the $c_{n,m}$ coefficients are numerical constants, but the
overall constant $\tilde{A}$ may depend on $B$. The density is obtained
from the residue of the resolvent at infinity 
\begin{equation}
U=\tilde{A}\frac{\sqrt{B}}{\pi}\sum_{m=0}^{\infty}c_{0,m}B^{-m}.
\end{equation}
For the Laplace transform study of its analytic properties leads to the following ansatz
\begin{equation}
\hat{R}(s)=\frac{A}{\sqrt{s}}\frac{\Gamma(1+s)}{\Gamma(\frac{1}{2}+s)}\left(\frac{1}{s+\frac{1}{2}}+\frac{1}{Bs}\sum_{n,m=0}^{\infty}\frac{Q_{n,m}}{B^{n+m}s^{n}}\right)\quad;\qquad A=\frac{e^{B}\sqrt{\pi}}{2\sqrt{2}}
\end{equation}
with constant coefficients $Q_{n,m}$.

The second seminal observation was that by re-expanding $R(\theta)$ and performing the
Laplace transform, the result can be compared to $\hat R(s)$. This gives $\tilde A = A$, and by matching
the two parametrizations, all the unknown numerical coefficients $Q_{n,m}$ and $c_{n,m}$ can be
determined completely algebraically in a recursive manner. This method was used by Volin
to calculate the perturbative coefficients for generic O$(N)$ models. The calculations in the
particular case of the O$(4)$ model take a simpler form.

\subsection{O$(4)$ Volin equations}

To describe the algebraic recursion equations for the O$(4)$ model we have to introduce a few
definitions.

The input parameters of the recursion are the coefficients $a_n$ of the expansion
\begin{equation}
\sum_{n=0}^\infty a_n x^n=g(x)=\frac{{\rm e}^{ax}}{1+x}\,\frac{\Gamma^2(1+x/2)}
{\Gamma(1+x)},\qquad a=\ln2,\qquad a_{-1}=0,  
\label{gx}
\end{equation}
and the rational constants
\begin{equation}
E_{n,k}=(-1)^k\frac{1}{2^k k!} \prod_{\ell=1}^k\left[n^2-\left(\frac{2\ell-1}{2}
\right)^2\right], \qquad E_{n,0}=1.
\end{equation}
Using the Gamma function representation
\begin{equation}
\Gamma(1+z)=\exp\left\{-\gamma z+\sum_{k=2}^\infty\frac{\zeta_k}{k}(-z)^k\right\}
\label{repGamma}
\end{equation}
we can re-express the basic input function (\ref{gx}) in terms of Dirichlet's
eta coefficients defined by
\begin{equation}
\eta_k=\sum_{n=1}^\infty (-1)^{n-1}\frac{1}{n^k}.
\end{equation}
They are simply related to the zeta coefficients by
\begin{equation}
\eta_k=(1-2^{1-k})\zeta_k,\quad k\geq2,\qquad\quad \eta_1=\ln2.
\label{etak}
\end{equation}
Also using the expansion
\begin{equation}
\ln (1+x)=\sum_{k=1}^\infty (-1)^{k-1} \frac{x^k}{k}  
\end{equation}
$g(x)$ can be recast in the form
\begin{equation}
\sum_{n=0}^\infty a_n x^n=g(x)=\exp\left\{\sum_{k=1}^\infty \frac{(-1)^{k-1}}{k}
(\eta_k-1) x^k\right\}.
\end{equation}
For later convenience we introduce the new two-index variable $p_{n,m}$ by
\begin{equation}
Q_{k,m}=\frac{p_{m+k,k}}{2^{m+2k+1}}.
\end{equation}
We now (recursively in $m$) find the unknowns $c_{n,m}$ ($m=0,1,\dots$,
$n=0,1,\dots$) and $p_{a,b}$ ($a=0,1,\dots$, $b=0,1,\dots,a$), starting from
\begin{equation}
c_{n,0}=a_n.  
\end{equation}
After having completed the recursion up to step $m-1$, we first have to solve
the set of equations $r=1,\dots,m$
\begin{equation}
\sum_{k=r}^m E_{k-r,k}\,c_{k-r,m-k}=\frac{1}{2^m}\sum_{k=r}^m (a_{k-r}+a_{k-r-1})
p_{m-1,k-1}  
\label{rec1}
\end{equation}
for $p_{m-1,k-1}$, $k=1,\dots,m$ and then use this solution to express
\begin{equation}
c_{n,m}=-\sum_{k=1}^m E_{n+k,k}\,c_{n+k,m-k}+\frac{1}{2^m}\sum_{k=1}^m
(a_{n+k}+a_{n+k-1})p_{m-1,k-1}  
\label{rec2}  
\end{equation}
$n=0,1,\dots$
To calculate the energy density $W(B)$ and the particle density $U(B)$ we only
need $c_{0,m}$ and 
\begin{equation}
W_n=\sum_{k=0}^n p_{n,k}.  
\end{equation}
Explicitly,
\begin{equation}
W(B)=\frac{1}{2\pi}\int_{-B}^B{\rm d}\theta\, \chi(\theta)\cosh\theta=
\frac{{\rm e}^{2B}}{16}\epsilon(B),  
\qquad \epsilon(B)=1+2\sum_{n=1}^\infty\frac{W_{n-1}}{(2B)^n},
\end{equation}
\begin{equation}
U(B)=\frac{1}{2\pi}\int_{-B}^B{\rm d}\theta\, \chi(\theta)=
{\rm e}^B\,\sqrt{\frac{B}{8\pi}}\rho(B),  
\qquad \rho(B)=1+\sum_{n=1}^\infty\frac{c_{0,n}}{B^n}.
\end{equation}
Later we will study the expansion of the generalized charges
\begin{equation}
\begin{split}
W_p&(B)=\frac{1}{2\pi}\int_{-B}^B{\rm d}\theta\, \chi(\theta)\cosh p\theta=
\frac{{\rm e}^{pB}}{4\pi}\hat R(p/2)\\
&=\frac{{\rm e}^{(p+1)B}}{4\pi\sqrt{p}}\,2^p\,\frac{\Gamma^2(1+p/2)}
{\Gamma(2+p)}\,\epsilon_p,  
\end{split}
\label{Wpdef}
\end{equation}
where
\begin{equation}
\epsilon_p(z)=1+\frac{p+1}{p}\sum_{n=1}^\infty \frac{H_n(p)}{(pz)^n},
\qquad\quad z=2B,
\end{equation}
and
\begin{equation}
H_n(p)=\sum_{k=1}^n p_{n-1,n-k}\,p^k.
\end{equation}
For the original energy density
\begin{equation}
W(B)=W_1(B),\qquad\quad \epsilon(B)=\epsilon_1(B), \qquad\quad H_n(1)=W_{n-1}.  
\end{equation}
An alternative form of $\epsilon_p(z)$, in terms of the variable $q=1/p$, is
\begin{equation}
\epsilon_p(z)=1+(1+q)\sum_{n=1}^\infty \frac{E_n(q)}{z^n},\qquad\quad
E_n(q)=\sum_{k=0}^{n-1} p_{n-1,k} q^k.  
\label{alter}
\end{equation}

The input parameters $a_n$ are polynomial expressions (with rational
coefficients) in $\zeta_2$, $\zeta_3,\dots$ and $a=\ln2$ of degree $n$
($a=\ln2$ counts as degree 1). Previously we observed that $W_n$ and
$c_{0,m}$ do not depend on the even zeta functions $\zeta_2$, $\zeta_4,\dots$
Since they are proportional to powers of $\pi$, we call this property
$\pi$-independence. We now observed that not only the sum $W_n$ but all
$p_{a,b}$ coefficients individually are $\pi$-independent. However, the even
zeta functions do not completely disappear: the variables $c_{n,m}$, $n\geq1$
do depend on them.

In \cite{Abbott:2020mba,Abbott:2020qnl} we studied the solution of the recursive equations (\ref{rec1}) and (\ref{rec2})
with high precision numerics, up to $2000^{\rm th}$ order, studied the resurgence properties of the
asymptotic expansions, and compared the Borel resummation of the asymptotic series for
both $U(B)$ and $W(B)$ to the results obtained from the numerical solution of the original
TBA equation. In this paper we reproduce many of those previous results analytically. We
hope that our methods may be used later in related similar models too.

\section{Summary}

In this section we summarize the analytic solution of the recursive equations
and study the resurgence properties of the asymptotic expansions analytically.
We not only confirm our previous findings but also discover some new
properties of the asymptotic series.

\subsection{Properties of the $\varepsilon_p(z)$ series}

The second set of Volin's equations, (\ref{rec2}), can be solved for $c_{n,m}$
in closed form and using this solution in (\ref{rec1})
we can obtain closed equations for the $p_{a,b}$ type variables. The details of
the derivation are given in appendix \ref{appA}.
The simplified equations are of the form
\begin{equation}
{\cal L}^{(M)}_r={\cal R}^{(M)}_r,\qquad M=1,2,\dots,\qquad r=1,\dots,M,
\label{LReqs}
\end{equation}
where
\begin{equation}
\begin{split}  
{\cal R}^{(M)}_r&=\sum_{k=0}^{M-r}(a_k+a_{k-1})p_{M-1,k+r-1}\\
&=a_{M-r}p_{M-1,M-1}+\sum_{k=0}^{M-r-1} a_k[p_{M-1,k+r-1}+p_{M-1,k+r}]  
\end{split}
\end{equation}
and
\begin{equation}
{\cal L}^{(M)}_r=\sum_{n=0}^{M-r} X_{r,n} Y^{(M-r)}_n.
\end{equation}
Here $X_{r,n}$ are $M$-independent rational coefficients:
\begin{equation}
X_{r,n}=2^r(-2)^n\sum_{p=0}^n(-1)^p E_{p,p+r} E_{p+1,n-p}  
\label{Xrn}
\end{equation}
and the symbols $Y^{(M)}_n$ are
\begin{equation}
Y^{(M)}_n=
\sum_{k=1}^{M-n}(a_{n+k}+a_{n+k-1})\,p_{M-n-1,k-1},\quad n=0,\dots,M-1,
\qquad Y^{(M)}_M=a_M.  
\end{equation}
The equations (\ref{LReqs}) can be solved recursively in $M$. 

In appendix \ref{appB}, using this simplified form of Volin's equations
we prove the existence of the asymptotic (ASY) expansions
\begin{equation}
\frac{p_{N,N-j}}{\Gamma(N+1)}=\frac{1}{\pi}\left\{
\beta^{(j)}+\frac{\alpha^{(j)}_0}{N}+\frac{\alpha^{(j)}_1}{N(N-1)}  
+\frac{\alpha^{(j)}_2}{N(N-1)(N-2)}+\dots\right\}\qquad j=0,1,\dots  
\label{ansatz0}
\end{equation}
Here the coefficients are given by generating functions:
\begin{equation}
\sum_{j=0}^\infty \beta^{(j)} p^{j+1} =u(p)  
\qquad\qquad \sum_{j=0}^\infty \alpha^{(j)}_m p^{j+1} =u(p)A_m(p) \ \ \
m=0,1,\dots  
\end{equation} 
where
\begin{equation}
u(p)=\frac{p}{1+p}\,\frac{g(-p)}{g(p)}
\qquad\qquad A_m(p)=(p-1)\sum_{k=0}^m p_{m,k}(-1)^k p^{m-k}    
\end{equation}

These results can be used to study the $\epsilon_p(z)$ asymptotic series and
related alien derivatives\footnote{The definitions related to alien derivatives
are summarized in appendix \ref{appF}.}.  

First we note that
\begin{equation}
u(x)=\frac{x}{(1-x)2^{2x}}\,\frac{\Gamma(1+x)}
{\Gamma(1-x)}\,\frac{\Gamma^2(1-x/2)}{\Gamma^2(1+x/2)}    
\label{uxdef}
\end{equation}
and using the representation of the Gamma function given by (\ref{repGamma})
we see that we have $\pi$-independence for the leading terms $\beta^{(j)}$
and also for $\alpha^{(j)}_m$ (at least for the first few $m$ values, for which
the $\pi$-indepence of $p_{m,k}$ is known by explicit construction).
We also note that $u(x)$ can alternatively be written as
\begin{equation}
u(x)=\frac{x}{2^{2x}}\,\frac{\Gamma(1+x)}
{\Gamma(2-x)}\,\frac{\Gamma^2(1-x/2)}{\Gamma^2(1+x/2)},    
\end{equation}
hence for $\vert x\vert<2$ the only singular point of $u(x)$ is $x=-1$:
\begin{equation}
(x\approx -1):\quad u(x)\approx -\frac{1}{2}\frac{1}{1+x}=-\frac{1}{2}
\sum_{j=0}^\infty(-x)^j.  
\end{equation}
Therefore for large $j$
\begin{equation}
\beta^{(j)}\approx \frac{1}{2}(-1)^j
\qquad\qquad \alpha^{(j)}_m\approx \frac{1}{2}A_m(-1)(-1)^j
\qquad\qquad A_m(-1)=2(-1)^{m+1} W_m.
\label{largej}
\end{equation}

The central object of study is
\begin{equation}
\epsilon_p(z)=1+\frac{p+1}{p}\sum_{m=0}^\infty\frac{H_{m+1}(p)}{(pz)^{m+1}}.  
\end{equation}
For $-p$ we have
\begin{equation}
\epsilon_{-p}(z)=1+\sum_{m=0}^\infty\frac{A_m(p)}{(pz)^{m+1}}.  
\end{equation}
We can write alternatively
\begin{equation}
\epsilon_p(z)=1+\frac{p+1}{p}\sum_{m=0}^\infty\frac{\Gamma(m+1)
c^{(\epsilon_p)}_m}{z^{m+1}\,p^{m+1}},\qquad\quad
c^{(\epsilon_p)}_m=\frac{H_{m+1}(p)}{\Gamma(m+1)}.
\end{equation}
These expansion coefficients define the Borel space function
\begin{equation}
B(x)=\sum_{m=0}^\infty c^{(\epsilon_p)}_m x^m\,.  
\end{equation}
The Borel transform of $\epsilon_p$ is given by
\begin{equation}
\hat\epsilon_p(t)=\frac{p+1}{p^2}B\left(\frac{t}{p}\right).  
\end{equation}

Now it is easy to establish the ASY
expansion of the Borel transform coefficients
(${\rm for}\ \vert p\vert <1$).
\begin{equation}
\begin{split}
c^{(\epsilon_p)}_M=\frac{H_{M+1}(p)}{\Gamma(M+1)}&=
\sum_{j=0}^M \frac{p_{M,M-j}}{\Gamma(M+1)}\,p^{j+1}\approx
\sum_{j=0}^\infty \frac{p^{j+1}}{\pi}\Big\{\beta^{(j)}+\sum_{m=0}^\infty
\frac{\alpha^{(j)}_m}{M_{[m+1]}}\Big\}\\
&=\frac{u(p)}{\pi}\Big\{1+\sum_{m=0}^\infty \frac{A_m(p)}{M_{[m+1]}}\Big\}.
\end{split}
\end{equation}
We have extended the upper limit of the summation from $M$ to $\infty$. This
is possible since the error we make by this extension is exponentially small
for $\vert p\vert<1$. 
Above we used the notation
\begin{equation}
M_{[r]}=M(M-1)(\dots)(M-r+1),\qquad\qquad M_{[1]}=M,\qquad M_{[0]}=1.  
\end{equation}

From here we can read off the singular part of the Borel space function $B(x)$:
\begin{equation}
B^{{\rm sing}}(x)=\frac{u(p)}{\pi}\left\{\frac{1}{1-x}-\ln(1-x)\sum_{m=0}^\infty
A_m(p)\frac{(x-1)^m}{m!}\right\}.
\end{equation}
The singular part of $\hat\epsilon_p(t)$ is
\begin{equation}
\hat\epsilon_p^{{\rm sing}}(t)=\frac{p+1}{p}
\frac{u(p)}{\pi}\left\{\frac{1}{p-t}-\ln(p-t)\sum_{m=0}^\infty
\frac{A_m(p)}{p^{m+1}}\frac{(t-p)^m}{m!}\right\}.
\end{equation}
Expressed in terms of alien derivatives we have
\begin{equation}
(\Delta_p\epsilon_p)(z)=-2i\frac{p+1}{p}u(p)\left\{
1+\sum_{m=0}^\infty \frac{A_m(p)}{(pz)^{m+1}}\right\}=-2i\frac{p+1}{p}u(p)
\epsilon_{-p}(z).  
\end{equation}
We can make it look simpler by introducing the rescaled
function\footnote{Note that $W_p= e^{(p+1)B}/(4\pi\sqrt{p}) \tilde{\epsilon_p}$.
See (\ref{Wpdef}).}  
\begin{equation}
\tilde\epsilon_p(z)=g(p)\epsilon_p(z).  
\end{equation}
In terms of the rescaled functions the alien derivative is simply
\begin{equation}
\Delta_p\tilde\epsilon_p=-2i\tilde\epsilon_{-p}.  
\end{equation}

We also note that since
\begin{equation}
\epsilon_{-1}(z)\equiv1,\qquad\qquad u(1)=1  
\end{equation}
in the $p\to1$ limit we have
\begin{equation}
\Delta_1\epsilon=-4i,  
\end{equation}
which is what we found earlier
numerically in \cite{Abbott:2020mba,Abbott:2020qnl}.
It means that the only singularity
of $\hat\epsilon(t)$ around $t=1$ is a simple pole with residue $-2/\pi$.

\subsection{\lq\lq Derivation'' of the naive ansatz}

Using (\ref{largej}) we see that the asymptotics if $j$ is also large
simplifies to
\begin{equation}
\frac{p_{N,N-j}}{\Gamma(N+1)}=\frac{(-1)^j}{2\pi}\left\{
1+\frac{A_0(-1)}{N}+\frac{A_1(-1)}{N(N-1)}  
+\frac{A_2(-1)}{N(N-1)(N-2)}+\dots\right\}  
\label{truej}
\end{equation}
The precise meaning of the above formula is that first we consider the large $N$
asymptotics of $p_{N,N-j}$ for fixed $j$ and next we let the fixed $j$ to be
large. If both $N$ and $j$ are very large but finite, the requirement is that
\begin{equation}
N\gg j\gg 1.
\label{requ}  
\end{equation}

We will now make the following bold assumption: (\ref{truej}) still holds if we
take $j=N-k$ with fixed $k$.  $N$ and $j$ are both large in this case but
the above requirement is not satisfied. Making this (illegal) substitution
we obtain
\begin{equation}
\frac{p_{N,k}}{\Gamma(N+1)}=\frac{(-1)^N}{2\pi}(-1)^k\left\{
1+\frac{\tilde R_0}{N}+\frac{\tilde R_1}{N(N-1)}  
+\frac{2\tilde R_2}{N(N-1)(N-2)}+\dots\right\}\qquad k=0,1,\dots  
\label{bold}
\end{equation}
with
\begin{equation}
\tilde R_m=\frac{2}{m!}(-1)^{m+1} W_m
\label{rela}
\end{equation}

We tried to check numerically if our bold assumption is correct. We started
from a somewhat more general ansatz:
\begin{equation}
\frac{p_{N,k}}{\Gamma(N+1)}=\frac{(-1)^N}{2\pi}(-1)^k\left\{
b^{(k)}+\frac{\tilde R^{(k)}_0}{N}+\frac{\tilde R^{(k)}_1}{N(N-1)}  
+\frac{2\tilde R^{(k)}_2}{N(N-1)(N-2)}+\dots\right\}\qquad k=0,1,\dots  
\label{ansatz}
\end{equation}
and found the first few coefficients numerically:
\begin{equation}
\begin{split}
b^{(k)}&=1\qquad k=0,1,\dots\\
\tilde R^{(0)}_0&=-\frac{3}{4}\qquad \tilde R^{(k)}_0=\tilde R_0=
-\frac{1}{2}=-2W_0 \qquad k=1,2,\dots\\
\tilde R^{(0)}_1&=\frac{49}{32}-\frac{3a}{2}\qquad \tilde R^{(1)}_1=\frac{27}{32}
-a\qquad \tilde R^{(k)}_1=\tilde R_1=\frac{9}{8}-a=2W_1 \qquad k=2,3,\dots\\
\tilde R^{(0)}_2&=-\frac{585}{256}+\frac{49a}{16}-\frac{3a^2}{2}\qquad
\tilde R^{(1)}_2=-\frac{45}{32}+\frac{27a}{16}-a^2\qquad
\tilde R^{(2)}_2=-\frac{531}{256}+\frac{9a}{4}-a^2\\
\tilde R^{(k)}_2&=\tilde R_2
=-\frac{57}{32}+\frac{9a}{4}-a^2=-W_2 \qquad k=3,4,\dots
\end{split}    
\end{equation}
We see that the leading term is correctly given by our bold assumption.
(\ref{bold}) is not completely correct but we found the following
interesting structure. There are irregular and regular coefficients:
\begin{equation}
{\rm irregular:}\quad \tilde R^{(k)}_m\quad k=0,\dots,m 
\qquad\qquad {\rm regular:}\quad \tilde R^{(k)}_m=\tilde R_m
=\frac{2}{m!}(-1)^{m+1} W_m\quad k\geq m+1. 
\label{irreg}
\end{equation}
We see that with the exception of a few irregular coefficients the generic
coefficients are correctly given by (\ref{bold}).

\subsection{The correct ansatz} 

Motivated by the naive \lq\lq derivation'' and based on our numerical studies
we now start from the ansatz (\ref{ansatz}). Unlike the case for our earlier
ansatz (\ref{ansatz0}) we cannot analytically prove the existence of the
asymptotic expansion (\ref{ansatz}). However, taking this form 
(with undetermined expansion coefficients $b^{(k)}$, $\tilde R^{(k)}$) for
granted, we can analytically prove the special structure we found numerically.

The starting point is (\ref{LReqs}) and using (\ref{ansatz}) and
comparing the asymptotic expansions of the two sides of the equations first
of all we can establish the generic structure
\begin{equation}
\left.\begin{split}
b^{(k)}&=b\\
\tilde R^{(k)}_m&=\tilde R_m\qquad k\geq m+1
\end{split}\right\}
{\rm \ \ with\ some\ generic\ } b,\tilde R_m.
\label{genericstr}
\end{equation}
Moreover, all irregular $\tilde R^{(k)}_m$ coefficients ($k=0,\dots,m$) can be
expressed linearly in terms of the generic parameters $b$, $\tilde R_m$.
In particular, up to NNNLO we find the result
\begin{equation}
\begin{split}
\tilde R^{(0)}_0&=\tilde R_0-p_{0,0}b,\\
\tilde R^{(0)}_1&=\tilde R_1-p_{0,0}\tilde R_0
+p_{1,0}b,\\
\tilde R^{(1)}_1&=\tilde R_1-p_{1,1}b,\\
\tilde R^{(0)}_2&=\tilde R_2-\frac{1}{2}p_{0,0}\tilde R_1
+\frac{1}{2}p_{1,0}\tilde R_0
-\frac{1}{2}p_{2,0}b,\\
\tilde R^{(1)}_2&=\tilde R_2-\frac{1}{2}p_{1,1}\tilde R_0+\frac{1}{2}p_{2,1}b,\\
\tilde R^{(2)}_2&=\tilde R_2-\frac{1}{2}p_{2,2}b.
\end{split}
\label{tilderesp}
\end{equation}
The proof of the generic structure (\ref{genericstr}) and the
technical details of the calculation leading to (\ref{tilderesp}) can be found
in appendix \ref{appC}.

The above findings suggest a product form of $\Delta_{-1}\epsilon_p$:
\begin{equation}
\Delta_{-1}\epsilon_p=iX\epsilon_p\qquad\qquad
X=b+\sum_{m=0}^\infty\frac{X_m}{z^{m+1}}.
\label{pr}
\end{equation}
The consequences of this relation are discussed in appendix \ref{appD}.
Writing out (\ref{pr}) explicitly we obtain
\begin{eqnarray}
\tilde R^{(k)}_m&=&\tilde R_m=\frac{(-1)^{m+1}}{m!} X_m,\quad k\geq m+1,
\label{comp2a}\\  
\tilde R^{(m)}_m&=&\tilde R_m-\frac{p_{m,m}}{m!} b,\label{comp2b}\\  
\tilde R^{(k)}_m&=&\tilde R_m+\frac{(-1)^{m+k+1}}{m!}\left\{
b p_{m,k}+\sum_{r=0}^{m-1-k} X_r p_{m-r-1,k}\right\},
\quad k\leq m-1.\label{comp2c}  
\end{eqnarray}
This gives the same structure as above and up to NNNLO explicitly reproduces
(\ref{tilderesp}) taking into account the identification (\ref{comp2a}).
The \lq\lq naive'' identification (\ref{rela}) gives $b=1$ and
$X_m=2W_m$, i.e. $X=\epsilon$. Finally we get
\begin{equation}
\Delta_{-1}\epsilon_p=i\epsilon\epsilon_p.  
\end{equation}
This is the generalization of the relation
\begin{equation}
\Delta_{-1}\epsilon=i\epsilon^2  
\end{equation}
found numerically in \cite{Abbott:2020mba,Abbott:2020qnl}.

\subsection{Asymptotic expansion of $\rho$}

From the inversion formula we obtain the coefficients of the Borel transform
of $\rho$:
\begin{equation}
c^{(\rho)}_N=\sum_{n=0}^{N+1}\omega_n \frac{Y^{(N+1)}_n}{\Gamma(N+1)},\qquad\qquad
\omega_n=(-2)^n E_{1,n}.  
\end{equation}
For the asymptotic expansion of $c^{(\rho)}_N$ up to NNNLO we need the
asymptotics of
\begin{equation*}
\frac{Y^{(N+1)}_n}{\Gamma(N+1)}\qquad\qquad {\rm for}\qquad\qquad n=0,1,2,3.
\end{equation*}
To calculate the above expansions the parameters
\begin{equation*}
b^{(k)},\ \tilde R^{(k)}_m\qquad m=0,1,2\qquad\qquad k=0,1,2,3  
\end{equation*}
are needed. We already expressed all these in terms of the generic coefficients.
The alien derivative of $\rho$ to this order is calculated in
appendix \ref{appE}. The result is
\begin{equation}
\Delta_{-1}\rho=i\left\{b+\frac{X_0+\tilde u_1 b}{z}  
+\frac{X_1+\tilde u_1 X_0+\tilde u_2 b}{z^2}  
+\frac{X_2+\tilde u_1 X_1+\tilde u_2 X_0+\tilde u_3 b}{z^3}+\dots\right\}.
\label{alienrho}
\end{equation}
Here we use the variables
\begin{equation}
\tilde u_M=2^M c_{0,M},
\end{equation}
which appear in the $\rho$ asymptotic series:
\begin{equation}
\rho=1+\sum_{M=1}^\infty \frac{c_{0,M}}{B^M}=\sum_{M=0}^\infty
\frac{\tilde u_M}{z^M}.
\end{equation}
The meaning of (\ref{alienrho}) is that (to this order)
\begin{equation}
\Delta_{-1}\rho=iX\rho=i\epsilon\rho.
\end{equation}

\section{Conclusion}

In this paper we investigated the perturbative expansion of the ground-state
energy $\epsilon$ and the generalized charges $\epsilon_{p}$ of
the O$(4)$ model in a magnetic field. We derived a closed system
of equations for these perturbative coefficients, which we solved
asymptotically by calculating the leading factorial growth together
with all the subsequent subfactorial behaviour. From this analytical
solution we could show that the asymptotic coefficients contain
only odd zeta numbers, a result similar to what was observed
before for the original perturbative coefficients by explicit calculation
\cite{Volin:2009wr,Volin:2010cq}. Our results also show that
the leading singularity of the Borel transform of $\epsilon_{p}$
is a logarithmic cut and the corresponding alien derivative resurges
to the same set of charges as 
\begin{equation}
\Delta_{p}\epsilon_{p}=-2i\frac{p+1}{p}u(p)\epsilon_{-p}
\end{equation}
with $u(p)$ an explicitly known function (\ref{uxdef}). This, in particular,
implies our previous finding $\Delta_{1}\epsilon=-4i$ for the energy
density \cite{Abbott:2020mba,Abbott:2020qnl}. We could also selfconsistently
determine the behaviour around the next singularity on the Borel plane,
which manifested itself as a logarithmic cut starting at $-1.$ The
corresponding alien derivative again showed resurgence in the form
\begin{equation}
\Delta_{-1}\epsilon_{p}=i\epsilon\epsilon_{p},
\end{equation}
 which is the generalization of what we found earlier numerically:
$\Delta_{-1}\epsilon=i\epsilon^{2}$. We extended this analysis also
for the density and showed that $\Delta_{-1}\rho=i\epsilon\rho$,
again conforming previous numerical observation. By this we completed
the investigations of the leading singularities. Our approach, however,
is quite general and is capable of investigating subsequent, exponentially
suppressed asymptotics, i.e. further cuts on the Borel plane. We are
planning to advance into this direction later. 

The recursive equations have their inputs from the asymptotic form
of the Laplace transform of the resolvent, $g(x)$, and the functions
$E_{n,m}$ playing a crucial role in moving between a function and
its Laplace transform, and which itself shows a kind of resurgence property
(\ref{EaN}). As the model-specific input comes only from $g(x)$ we analyzed
a toy model, where we replaced all $\eta_{n}$ (see (\ref{etak})) with $1$.
This was also motivated by the fact that numerically they are not
far from each other. We could solve this model explicitly, see
appendix~\ref{eta1}, and observed a similar resurgence pattern to what we found
for the real problem.
This sheds some light on their origins. This toy model then
can be systematically dressed up, by including corrections in 
$\eta_{n}-1$. In particular, the exact dependence on $\eta_1$
is found in appendix~\ref{eta1}. 

In the present paper we analysed the density and the generalized charges
as a function of the TBA parameter $B$. The change for the running
coupling is well understood and the composition formula for alien
derivatives \cite{composit} allows a direct translation \cite{Abbott:2020mba,Abbott:2020qnl}. 

So far we analyzed only the O$(4)$ model, but the technics we developed here can also
be extended for other O$(N)$ models and their supersymmetric generalizations as well as
for the principal chiral and Gross-Neveu models, all of which were analysed recently in
\cite{Marino:2021dzn} by directly investigating the integral equation.
They may be useful also for non-relativistic two dimensional
integrable theories relevant for statistical physics such
as \cite{Marino:2019fuy,Marino:2020dgc,Marino:2020ggm}.

\subsection*{Acknowledgments}

We thank Gerg\H{o} Nemes for useful discussions, the Isaac Newton
Institute programme 'Applicable resurgent asymptotics: towards a universal
theory', and the NKFIH Grant 134946 for support. 



%

\appendix

\section{Simplifying Volin's equations}
\label{appA}

\subsection{Properties of the $a_n$ coefficients}
  
The input $a_n$ parameters are obtained from the Taylor expansion of (\ref{gx}).
The first few coefficients are
\begin{equation}
a_0=1,\qquad\quad a_1=a-1,\qquad\quad a_2=1-a+\frac{a^2}{2}-\frac{\pi^2}{24}.  
\end{equation}
$g(x)$ can also be represented as
\begin{equation}
g(x)=2^{x+1}\,\frac{\Gamma^2(2+x/2)}{\Gamma(3+x)}\,\frac{1}{1+x/2}=
\frac{1}{2+x}+{\rm conv}_4(x),  
\end{equation}
where the second term in the last formula is some function having a 
power series expansion convergent for
$\vert x\vert<4$. This form implies that for large $n$
\begin{equation}
a_n\sim\frac{(-1)^n}{2^{n+1}}.  
\end{equation}

\subsection{$c_{n,m}$ equations}

For later purposes we note that
\begin{equation}
E_{1,k}=\frac{(-1)^k}{2^k}\omega_k,  
\end{equation}
where
\begin{equation}
\omega_k=\frac{1}{k!}\,\prod_{\ell=1-k}^k\left(\ell+\frac{1}{2}\right),\quad
k\geq1,\qquad\quad \omega_0=1.  
\end{equation}
We introduce the notation
\begin{equation}
c_{n,m}=w_{n,n+m}.  
\end{equation}
With this notation the $c_{n,m}$ equations for fixed $M\geq1$ and
$n=0,1,\dots M-1$ read
\begin{equation}
\sum_{k=n}^M E_{k,k-n}\,w_{k,M}=2^{n-M}Y^{(M)}_n,
\label{weqs}  
\end{equation}
where
\begin{equation}
Y^{(M)}_n=
\sum_{k=1}^{M-n}(a_{n+k}+a_{n+k-1})\,p_{M-n-1,k-1},\qquad\quad n=0,\dots,M-1.
\end{equation}
The unknowns in these equations are
\begin{equation}
w_{j,M},\qquad j=0,\dots,M-1
\end{equation}
since for the last component
\begin{equation}
w_{M,M}=c_{M,0}=a_M,  
\end{equation}
which is an input. However, we can formally extend (\ref{weqs}) for $n=M$ as
well, if we define
\begin{equation}
Y^{(M)}_M=a_M.  
\end{equation}
This way we have $M+1$ equations for $M+1$ unknowns and the equations can be
written in matrix form:
\begin{equation}
\sum_{k=0}^M{\cal M}_{n,k}\, w_{k,M}=2^{n-M}Y^{(M)}_n,
\end{equation}
where
\begin{equation}
{\cal M}_{n,k}=\left\{
\begin{split}
E_{k,k-n}\quad k&\geq n\\
0\qquad k&<n
\end{split}\right. \,.
\label{Mmatrix}
\end{equation}
The solution is given by the inverse matrix:
\begin{equation}
w_{a,M}=\sum_{n=0}^M{\cal M}^{-1}_{a,n}\,2^{n-M}\,Y^{(M)}_n.
\end{equation}
We are mostly interested in
\begin{equation}
c_{0,M}=w_{0,M}=\sum_{n=0}^M{\cal M}^{-1}_{0,n}\,2^{n-M}\,Y^{(M)}_n,
\end{equation}
because these are the ones appearing in the $B$-expansion of $\rho(B)$.

The inverse matrix is also triangular and we found the simple result
\begin{equation}
{\cal M}^{-1}_{a,n}=(-1)^{n-a} E_{a+1,n-a}, \qquad (n\geq a).  
\label{Minverse}
\end{equation}
In particular,
\begin{equation}
{\cal M}^{-1}_{0,n}=(-1)^n E_{1,n}=\frac{\omega_n}{2^n}.  
\end{equation}

For some purposes it is useful to
express the variables $w_{a,m}$ from the second set of Volin equations and
if we use
\begin{equation}
w_{a,M}=\sum_{n=a}^M(-1)^{n+a} E_{a+1,n-a} 2^{n-M} Y^{(M)}_n
\end{equation}
we can obtain closed equations for the $p_{a,b}$ type variables, which
can be solved recursively in $M$. These equations are given by
(\ref{LReqs})-(\ref{Xrn}).

\section{Analytic proof of the asymptotic expansion (\ref{ansatz0})}
\label{appB}

\subsection{Some definitions, notations, relations}

The definition of the expansion coefficients of the basic input function
$g(x)$ is extended to negative indices as
\begin{equation}
a_{-n}=0,\qquad\quad n=1,2,\dots
\label{nega}  
\end{equation}
If a function ${\cal F}(M)$ has asymptotic series (ASY) for large $M$, the
coefficients of the individual terms in this expansion will be denoted by
${\cal F}^{[r]}$:
\begin{equation}
{\cal F}(M)=\sum_{r=0}^\infty\frac{{\cal F}^{[r]}}{M_{[r]}}=
{\cal F}^{[0]}+\frac{{\cal F}^{[1]}}{M}+\frac{{\cal F}^{[2]}}{M(M-1)}    
+\frac{{\cal F}^{[3]}}{M(M-1)(M-2)}+\dots    
\end{equation}
In this appendix $M$ is always assumed to be asymptotically large. For
simplicity, the $M$ dependence will not always be explicit. For example, we will
use the notations
\begin{equation}
\Omega^{(k)}=\frac{p_{M,M-k}}{\Gamma(M+1)},\qquad\quad k=0,1,\dots  
\end{equation}
and
\begin{equation}
\bar {\cal R}_j=\frac{{\cal R}^{(M+1)}_{M+1-j}}{\Gamma(M+1)},\qquad\qquad
\bar {\cal L}_j=\frac{{\cal L}^{(M+1)}_{M+1-j}}{\Gamma(M+1)},\qquad\quad
j=0,1,\dots
\end{equation}

This appendix is about the analytic proof of the following two statements.

\underline{Statement 1}

$\Omega^{(j)}$ ($j=0,1,\dots$) has ASY expansion
\begin{equation}
\begin{split}  
\frac{p_{M,M-j}}{\Gamma(M+1)}&=\frac{1}{\pi}\Big\{\beta^{(j)}
+\frac{\alpha^{(j)}_0}{M}+\frac{\alpha^{(j)}_1}{M(M-1)}  
+\frac{\alpha^{(j)}_2}{M(M-1)(M-2)}+\dots\Big\}\\
&=\frac{1}{\pi}\Big\{\beta^{(j)}+\sum_{m=0}^\infty\frac{\alpha^{(j)}_m}{M_{[m+1]}}
\Big\}.  
\end{split}
\end{equation}
Here we state the existence of the ASY expansion and define the expansion
coefficients.


\underline{Statement 2}

The expansion coefficients are given with the help of generating functions
as follows.
\begin{equation}
\sum_{j=0}^\infty \beta^{(j)}\,p^{j+1}=u(p),\qquad\qquad
\sum_{j=0}^\infty \alpha^{(j)}_m\,p^{j+1}=u(p)A_m(p),\qquad\qquad m=0,1,\dots
\end{equation}
Here
\begin{equation}
u(p)=\frac{p}{p+1}\,\frac{g(-p)}{g(p)}.
\end{equation}

\subsection{$E_{a,N}$ resurgence}

For $N>a$ an alternative representation of $E_{a,N}$ is
\begin{equation}
E_{a,N}=(-1)^a\frac{\Gamma(2N-2a+1)\Gamma(2N+2a+1)}{2^{5N}\Gamma(N+1)
\Gamma(N-a+1)\Gamma(N+a+1)}.  
\end{equation}
The following combination has ASY expansion for large $N$, which can be
calculated using the Stirling series.
\begin{equation}
\begin{split}
\frac{\pi}{\Gamma(N)}&(-1)^a2^NE_{a,N}=\pi
\frac{\Gamma(2N-2a+1)\Gamma(2N+2a+1)}{2^{4N}\Gamma(N)\Gamma(N+1)
\Gamma(N-a+1)\Gamma(N+a+1)}\\
&=1+\sum_{m=1}^\infty\frac{1}{m!}\prod_{\ell=1}^m\frac{a^2-(\ell-1/2)^2}{N-\ell}
=1+\sum_{m=1}^\infty\frac{(-2)^mE_{a,m}}{(N-1)(\dots)(N-m)}.
\end{split}    
\end{equation}
For later purpose we write the above expansion as
\begin{equation}
2^NE_{a,N}=\frac{(-1)^a\Gamma(N)}{\pi}\left\{1+\sum_{m=1}^\infty
\frac{(-2)^mE_{a,m}}{(N-1)(\dots)(N-m)}\right\},  
\label{EaN}
\end{equation}
which can be interpreted as a kind of \lq\lq resurgence'' of the function
$E_{a,N}$.

\subsection{Proof of the statements}

We start from (\ref{LReqs}) in the form
\begin{equation}
\bar{\cal L}_j=\bar{\cal R}_j,\qquad\qquad j=0,1,\dots  
\label{LjRj}
\end{equation}
The two sides of the equations are
\begin{equation}
\bar{\cal R}_j=\sum_{k=0}^j(a_k+a_{k-1})\Omega^{(j-k)}  
\label{barRj}
\end{equation}
and
\begin{equation}
\bar{\cal L}_j=\sum_{n=0}^j\frac{X_{M+1-j,n}}{\Gamma(M+1)}\, Y^{(j)}_n.
\label{barLj}
\end{equation}
First we consider the large $M$ asymptotics of the left hand side.
\begin{equation}
\begin{split}
&\frac{X_{M+1-j,n}}{\Gamma(M+1)}=2^{M+1-j}(-2)^n\sum_{p=0}^n(-1)^p\frac{
E_{p,p+M+1-j}}{\Gamma(M+1)}\,E_{p+1,n-p}\\
&=(-2)^n\sum_{p=0}^n 2^{-p} E_{p+1,n-p}\frac{1}{\pi}\frac{1}{M_{[j-p]}}\Big\{
1+\sum_{m=1}^\infty \frac{(-2)^m E_{p,m}}{(p+M-j)(\dots)(p+M-j+1-m)}\Big\}\\
&=\frac{(-2)^n}{\pi}\sum_{p=0}^n 2^{-p} E_{p+1,n-p}\sum_{m=0}^\infty\frac{(-2)^m
E_{p,m}}{M_{[j-p+m]}}.  
\end{split}    
\end{equation}
The $r^{{\rm th}}$ component of the ASY expansion is given by the
$m=r+p-j$ term (which is nonzero if $n\geq p\geq j-r$):  
\begin{equation}
\left(\frac{X_{M+1-j,n}}{\Gamma(M+1)}\right)^{[r]}=
\frac{(-2)^n}{\pi}\sum_{p={\rm Max}(0,j-r)}^n 2^{-p} E_{p+1,n-p}(-2)^{r+p-j}E_{p,r+p-j}.
\label{Xr}
\end{equation}
If $j\geq r$ we can recognize here the triangular matrix (\ref{Mmatrix})
and its inverse (\ref{Minverse}) and the above can be written as
\begin{equation}
\frac{(-1)^{r-j}}{\pi}2^{r-j+n}\sum_{p=j-r}^n{\cal M}^{-1}_{p,n}\,{\cal M}_{j-r,p}.
\end{equation}
Since
\begin{equation}
{\cal M}^{-1}_{p,n}=0\qquad{\rm if\ }p>n\qquad\qquad{\rm and}\qquad\qquad
{\cal M}_{j-r,p}=0\qquad{\rm if\ }p<j-r,  
\end{equation}
we can extend the summation limits to $\sum_{p=0}^M$ and the summation becomes
the matrix product of the matrix and its inverse and the result is simply the
Kronecker delta
\begin{equation}
\delta_{n,j-r}.  
\end{equation}
Using this in (\ref{barLj}) we get
\begin{equation}
\bar{\cal L}_j^{[r]}=\frac{(-1)^{j-r}}{\pi}\, Y^{(j)}_{j-r}.
\end{equation}
Explicitly,
\begin{equation}
\begin{split}
\bar{\cal L}_j^{[0]}&=\frac{(-1)^j}{\pi}a_j,\\    
\bar{\cal L}_j^{[r]}&=\frac{(-1)^{j-r}}{\pi}\sum_{k=0}^{r-1}(a_{j-r+k+1}
+a_{j-r+k})p_{r-1,k},\qquad\quad j\geq r\geq1.    
\end{split}    
\end{equation}
For $j<r$ (\ref{Xr}) gives
\begin{equation}
\left(\frac{X_{M+1-j,n}}{\Gamma(M+1)}\right)^{[r]}=
\frac{(-2)^{n+r-j}}{\pi}
\sum_{p=0}^n (-1)^p E_{p+1,n-p}E_{p,r+p-j}=\frac{(-1)^{r-j}}{\pi}\,X_{r-j,n}.
\end{equation}
Using this in (\ref{barLj}) we get
\begin{equation}
\bar{\cal L}_j^{[r]}=\frac{(-1)^{r-j}}{\pi}\sum_{n=0}^j X_{r-j,n}Y^{(j)}_n
=\frac{(-1)^{r-j}}{\pi}{\cal L}^{(r)}_{r-j}.
\end{equation}
Using (\ref{LReqs}) once more in this case we have
\begin{equation}
\begin{split}
\bar{\cal L}_j^{[r]}&=\frac{(-1)^{j-r}}{\pi}{\cal R}^{(r)}_{r-j}=\frac{(-1)^{j-r}}
{\pi}\sum_{k=0}^{j}(a_k+a_{k-1})p_{r-1,k+r-j-1}\\
&=\frac{(-1)^{j-r}}{\pi}\sum_{k=r-j-1}^{r-1}(a_{k+j-r+1}+a_{k+j-r})p_{r-1,k}.
\end{split}    
\end{equation}
Finally, using the convention (\ref{nega}) the summation limits can be extended:
\begin{equation}
\bar{\cal L}_j^{[r]}=\frac{(-1)^{j-r}}{\pi}\sum_{k=0}^{r-1}(a_{j-r+k+1}
+a_{j-r+k})p_{r-1,k},   
\end{equation}
which is miraculously the same as the formula for $j\geq r$!

To summarize, the final result of the calculation is (for any $j=0,1,\dots$) 
\begin{equation}
\begin{split}
\bar{\cal L}_j^{[0]}&=\frac{(-1)^j}{\pi}a_j,\\    
\bar{\cal L}_j^{[r]}&=\frac{(-1)^{j-r}}{\pi}\sum_{k=0}^{r-1}(a_{j-r+k+1}
+a_{j-r+k})p_{r-1,k},\qquad\quad r\geq1.    
\end{split}    
\end{equation}

This way we have proven by explicit calculation that the LHS of (\ref{LjRj})
has ASY expansion. But then also the RHS has ASY expansion. In particular
\begin{equation}
\Omega^{(0)}=\bar{\cal R}_0=\bar{\cal L}_0  
\end{equation}
has ASY expansion, then
\begin{equation}
\Omega^{(1)}=-(1+a_1)\Omega^{(0)}+\bar{\cal L}_1  
\end{equation}
has ASY expansion, and so on: we can prove recursively that all
\begin{equation}
\Omega^{(j)}=-\sum_{k=1}^j(a_k+a_{k-1})\Omega^{(j-k)}+\bar{\cal L}_j  
\end{equation}
have ASY expansion. Thus we have proven Statement 1.

The proof of Statement 2 is now easy. We calculate the infinite sums
\begin{equation}
\sum_{j=0}^\infty p^j\bar{\cal L}_j^{[0]}=\frac{1}{\pi}\sum_{j=0}^\infty(-p)^ja_j=
\frac{g(-p)}{\pi}  
\end{equation}
and (for $r\geq1$)
\begin{equation}
\begin{split}  
\sum_{j=0}^\infty p^j\bar{\cal L}_j^{[r]}&=\frac{(-1)^r}{\pi}\sum_{j=0}^\infty(-p)^j
\sum_{k=0}^{r-1}(a_{k+j-r+1}+a_{k+j-r})p_{r-1,k}\\
&=\frac{(-1)^r}{\pi}\sum_{k=0}^{r-1}p_{r-1,k}\sum_{i=0}^\infty a_i\big[(-p)^{i+r-k-1}
+(-p)^{i+r-k}\big]\\  
&=\frac{(p-1)g(-p)}{\pi}\sum_{k=0}^{r-1}p_{r-1,k}(-1)^k p^{r-1-k}=
\frac{g(-p)}{\pi}A_{r-1}(p).  
\end{split}
\end{equation}

For the RHS 
\begin{equation}
\Omega^{(j)[r]}=\frac{1}{\pi}\alpha^{(j)}_{r-1},  
\end{equation}
where we have temporarily introduced the short hand notation
\begin{equation}
\alpha^{(j)}_{-1}=\beta^{(j)}.
\end{equation}
The infinite summation for the RHS gives
\begin{equation}
\sum_{j=0}^\infty p^j\bar{\cal R}_j^{[r]}=\sum_{j=0}^\infty p^j\sum_{k=0}^j
(a_k+a_{k-1})\frac{1}{\pi}\alpha^{(j-k)}_{r-1}=\frac{1+p}{p}\frac{g(p)}{\pi}
\sum_{s=0}^\infty p^{s+1}\alpha^{(s)}_{r-1}.
\end{equation}
Comparing the infinite sums at the two sides we obtain
\begin{equation}
\begin{split}
\sum_{s=0}^\infty p^{s+1}\beta^{(s)}&=\frac{p}{1+p}\frac{g(-p)}{g(p)}=u(p),\\
\sum_{s=0}^\infty p^{s+1}\alpha^{(s)}_{r-1}&=\frac{p}{1+p}\frac{g(-p)}{g(p)}
A_{r-1}(p)=u(p)A_{r-1}(p),\qquad r\geq1.
\end{split}    
\end{equation}
This is Statement 2.

\section{Consistency of the conjectured asymptotics}
\label{appC}

In this appendix our starting point is the ansatz (\ref{ansatz}). However,
for technical reasons we will use its (equivalent) index-shifted version:
\begin{equation}
\frac{p_{M-1,k}}{\Gamma(M)}=\frac{(-1)^M}{2\pi}(-1)^{k+1}\left\{
b^{(k)}+\frac{R^{(k)}_0}{M}+\frac{R^{(k)}_1}{M(M-1)}  
+\frac{2R^{(k)}_2}{M(M-1)(M-2)}+\dots\right\}\qquad k=0,1,\dots  
\label{ansatz1}
\end{equation}
This asymptotic expansion is assumed valid for fixed $k=0,1,\dots$ in the
large $M$ limit.
Using this ansatz we can calculate the asymptotic expansion of both sides
of the Volin equations (\ref{LReqs}).

The large $M$ asymptotics of $Y^{(M)}_n$ can be written as
\begin{equation}
\begin{split}  
\frac{Y^{(M)}_n}{\Gamma(M)}
\approx &\frac{(-1)^{M-1}(-1)^n}{2\pi(M-1)(\cdots)(M-n)}
\sum_{k=0}^{M-n-1}(a_{n+k+1}+a_{n+k})(-1)^k\\  
\Big\{&b^{(k)}+\frac{R^{(k)}_{0}}{M-n}
+\frac{R^{(k)}_1}{(M-n)(M-n-1)}+\dots\Big\}.\\
\end{split}
\label{YnASY1}
\end{equation}
(\ref{ansatz1}) is only valid for $k\ll M$, but for large $k$ the coefficients
$a_{n+k}$ are exponentially small and the asymptotic expansion in powers of
$1/M$ is not affected by this condition. Using the same reasoning, we can
extend the $k$-summation to infinity:
\begin{equation}
\begin{split}  
\frac{Y^{(M)}_n}{\Gamma(M)}
\approx &\frac{(-1)^{M-1}(-1)^n}{2\pi(M-1)(\cdots)(M-n)}
\sum_{k=0}^\infty (a_{n+k+1}+a_{n+k})(-1)^k\\  
\Big\{&b^{(k)}+\frac{R^{(k)}_{0}}{M-n}
+\frac{R^{(k)}_1}{(M-n)(M-n-1)}+\dots\Big\}.\\
\end{split}
\label{YnASY2}
\end{equation}
We work up to NNNLO and rearrange the summation as follows.
\begin{equation}
\begin{split}
\frac{Y^{(M)}_n}{\Gamma(M)}&=\frac{(-1)^{M-n}}{2\pi(M-1)(\dots)(M-n)}\Big\{
-a_n\Big[b^{(0)}+\frac{R^{(0)}_0}{M-n}\\
&+\frac{R^{(0)}_1}{(M-n)(M-n-1)}
+\frac{2R^{(0)}_2}{(M-n)(M-n-1)(M-n-2)}\Big]\\  
&+\sum_{k=1}^\infty
a_{n+k}(-1)^k\Big[b^{(k-1)}-b^{(k)}+\frac{R^{(k-1)}_0-R^{(k)}_0}{M-n}\\
&+\frac{R^{(k-1)}_1-R^{(k)}_1}{(M-n)(M-n-1)}
+\frac{2[R^{(k-1)}_2-R^{(k)}_2]}{(M-n)(M-n-1)(M-n-2)}\Big]+\dots\Big\}.  
\label{ASYYMn}
\end{split}    
\end{equation}
and similarly
\begin{equation}
\begin{split}
\frac{Y^{(M-r)}_n}{\Gamma(M)}&=\frac{(-1)^{M-n-r}}{2\pi(M-1)(\dots)(M-n-r)}\Big\{
-a_n\Big[b^{(0)}+\frac{R^{(0)}_0}{M-n-r}\\
&+\frac{R^{(0)}_1}{(M-n-r)(M-n-r-1)}
+\frac{2R^{(0)}_2}{(M-n-r)(M-n-r-1)(M-n-r-2)}\Big]\\  
&+\sum_{k=1}^\infty
a_{n+k}(-1)^k\Big[b^{(k-1)}-b^{(k)}+\frac{R^{(k-1)}_0-R^{(k)}_0}{M-n-r}\\
&+\frac{R^{(k-1)}_1-R^{(k)}_1}{(M-n-r)(M-n-r-1)}
+\frac{2[R^{(k-1)}_2-R^{(k)}_2]}{(M-n-r)(M-n-r-1)(M-n-r-2)}\Big]+\dots\Big\}.  
\end{split}    
\end{equation}
Proceeding analogously for the RHS, up to NNNLO  we find
\begin{equation}
\begin{split}
\frac{{\cal R}^{(M)}_r}{\Gamma(M)}&=\frac{(-1)^r (-1)^M}{2\pi}\sum_{k=0}^\infty
a_k(-1)^k\Big\{b^{(k+r-1)}-b^{(k+r)}+\frac{R^{(k+r-1)}_0-R^{(k+r)}_0}{M}\\
&+\frac{R^{(k+r-1)}_1-R^{(k+r)}_1}{M(M-1)}
+\frac{2[R^{(k+r-1)}_2-R^{(k+r)}_2]}{M(M-1)(M-2)}+\dots\Big\}.  
\end{split}    
\end{equation}

The crucial observation is that
\begin{equation}
\frac{Y^{(M-r)}_n}{\Gamma(M)}={\rm O}(M^{-n-r}) 
\end{equation}
and so
\begin{equation}
\frac{{\cal L}^{(M)}_r}{\Gamma(M)}={\rm O}(M^{-r}), 
\end{equation}
and consequently we must also have
\begin{equation}
\frac{{\cal R}^{(M)}_r}{\Gamma(M)}={\rm O}(M^{-r}). 
\end{equation}
But this is possible only if
\begin{equation}
b^{(k)}=b,\qquad k=0,1,\dots
\end{equation}
and
\begin{equation}
R^{(k)}_m=R_m,\qquad k\geq m+1.
\end{equation}
This is precisely the structure we observed numerically.

Using the above simplifications, to a given order in the asymptotic expansion,
both sides of the equations contain a finite number of nonzero summands only.
We work at NNNLO and we need the following expressions.
\begin{equation}
\begin{split}
\frac{{\cal R}^{(M)}_1}{\Gamma(M)}&=\frac{(-1)^{M-1}}{2\pi}
\Big\{\frac{R^{(0)}_0-R_0}{M}
+\frac{R^{(0)}_1-R^{(1)}_1+a_1[R_1-R^{(1)}_1]}{M(M-1)}\\
&+\frac{2[R^{(0)}_2-R^{(1)}_2]+2a_1[R^{(2)}_2-R^{(1)}_2]
+2a_2[R^{(2)}_2-R_2]}{M(M-1)(M-2)}+\dots\Big\},  
\end{split}    
\end{equation}
\begin{equation}
\frac{{\cal R}^{(M)}_2}{\Gamma(M)}=\frac{(-1)^M}{2\pi}
\Big\{\frac{R^{(1)}_1-R_1}{M(M-1)}
+\frac{2[R^{(1)}_2-R^{(2)}_2]+2a_1[R_2-R^{(2)}_2]}{M(M-1)(M-2)}+\dots\Big\},  
\end{equation}
\begin{equation}
\frac{{\cal R}^{(M)}_3}{\Gamma(M)}=\frac{(-1)^{M-1}}{2\pi}
\Big\{\frac{2[R^{(2)}_2-R_2]}{M(M-1)(M-2)}+\dots\Big\},  
\end{equation}
\begin{equation}
\begin{split}
&\frac{Y^{(M-1)}_0}{\Gamma(M)}=\frac{(-1)^{M}}{2\pi}\Big\{\frac{b}{M}
+\frac{b+R^{(0)}_0+a_1[R^{(0)}_0-R_0]}{M(M-1)}\\
&+\frac{R^{(0)}_0+R^{(0)}_1+a_1 [R^{(0)}_0-R_0]+a_1[R^{(0)}_1-R^{(1)}_1]
+a_2[R_1-R^{(1)}_1]}{M(M-1)(M-2)}+\dots\Big\},  
\end{split}    
\end{equation}
\begin{equation}
\frac{Y^{(M-1)}_1}{\Gamma(M)}=\frac{(-1)^{M}}{2\pi}\Big\{
\frac{-ba_1}{M(M-1)}
-\frac{2ba_1+a_1R^{(0)}_0+a_2[R^{(0)}_0-R_0]}{M(M-1)(M-2)}+\dots\Big\},  
\end{equation}
\begin{equation}
\frac{Y^{(M-1)}_2}{\Gamma(M)}=\frac{(-1)^{M}}{2\pi}\Big\{
\frac{ba_2}{M(M-1)(M-2)}+\dots\Big\},  
\end{equation}
\begin{equation}
\frac{Y^{(M-2)}_0}{\Gamma(M)}=\frac{(-1)^{M}}{2\pi}\Big\{
\frac{-b}{M(M-1)}
-\frac{2b+R^{(0)}_0+a_1[R^{(0)}_0-R_0]}{M(M-1)(M-2)}+\dots\Big\},  
\end{equation}
\begin{equation}
\frac{Y^{(M-2)}_1}{\Gamma(M)}=\frac{(-1)^{M}}{2\pi}\Big\{
\frac{ba_1}{M(M-1)(M-2)}+\dots\Big\},  
\end{equation}
\begin{equation}
\frac{Y^{(M-3)}_0}{\Gamma(M)}=\frac{(-1)^{M}}{2\pi}\Big\{
\frac{b}{M(M-1)(M-2)}+\dots\Big\}.  
\end{equation}

Working up to NNNLO order, we need the asymptotic expansion of the combinations
\begin{equation}
\begin{split}
{\cal L}^{(M)}_1&=X_{1,0}Y^{(M-1)}_0+X_{1,1}Y^{(M-1)}_1+X_{1,2}Y^{(M-1)}_2+\dots,\\
{\cal L}^{(M)}_2&=X_{2,0}Y^{(M-2)}_0+X_{2,1}Y^{(M-2)}_1+\dots,\\
{\cal L}^{(M)}_3&=X_{3,0}Y^{(M-3)}_0+\dots
\end{split}     
\end{equation}
Comparing these expansions to ${\cal R}^{(M)}_1$, ${\cal R}^{(M)}_2$, and
${\cal R}^{(M)}_3$ and matching 3, 2, and 1 asymptotic expansion coefficients
respectively, we can write down 6 equations. These can be used to express 6
leading coefficients,
$R^{(0)}_0$, $R^{(0)}_1$, $R^{(1)}_1$, $R^{(0)}_2$, $R^{(1)}_2$, and $R^{(2)}_2$,
in terms of the \lq\lq generic'' ones, $b$, $R_0$, $R_1$, and $R_2$.
The result is
\begin{equation}
\begin{split}
R^{(0)}_0&=R_0-\frac{b}{4},\\
R^{(0)}_1&=R_1-\frac{R_0}{4}+\left(\frac{1}{32}-\frac{a}{2}\right)b,\\
R^{(1)}_1&=R_1-\frac{9b}{32},\\
R^{(0)}_2&=R_2-\frac{R_1}{8}+\left(\frac{1}{64}-\frac{a}{4}\right)R_0
+\left(-\frac{3}{256}+\frac{a}{16}-\frac{a^2}{2}\right)b,\\
R^{(1)}_2&=R_2-\frac{9R_0}{64}+\left(\frac{3}{128}-\frac{9a}{16}\right)b,\\
R^{(2)}_2&=R_2-\frac{75}{256}b.
\label{untilderes}
\end{split}    
\end{equation}
Rewriting the above result in terms of the original
\lq\lq tilde'' variables appearing in (\ref{ansatz}) we obtain
(\ref{tilderesp}).

\section{Resurgence}
\label{appD}

The results (\ref{tilderesp}) show that the first few
small-index $p_{a,b}$ quantities appear in the expression of the first few
asymptotic coefficients and this structure strongly suggests a product
form for the $\Delta_{-1}$ alien derivative.

In this appendix we will use the alternative representation of $\epsilon_p(z)$
given by (\ref{alter}) and write the Taylor expansion coefficients of its
Borel transform as
\begin{equation}
c^{(\epsilon_p)}_n=(1+q)\frac{E_{n+1}(q)}{\Gamma(n+1)},\qquad\quad n=0,1,\dots  
\end{equation}
We will assume that $\vert q\vert<1$.
Using the ansatz (\ref{ansatz}) we can write down the asymptotic expansion of
the Borel transform coefficients:
\begin{equation}
c^{(\epsilon_p)}_n=(1+q)\frac{(-1)^n}{2\pi}\sum_{k=0}^\infty (-q)^k
\left\{ b^{(k)}
+\frac{\tilde R^{(k)}_0}{n}+\frac{\tilde R^{(k)}_1}{n(n-1)}
+\frac{2\tilde R^{(k)}_2}{n(n-1)(n-2)}
+\dots\right\}.
\end{equation}
Here, again, we extended the upper limit of the summation to infinity. Comparing
this expansion with the generic forms (\ref{ASYcn}) and (\ref{alien}) we get
\begin{eqnarray}
\pi\tilde B_o&=&(1+q)\sum_{k=0}^\infty (-q)^k b^{(k)},\\
\pi\tilde B_o\tilde q_m&=&(1+q)\sum_{k=0}^\infty (-q)^k \tilde R^{(k)}_m,
\qquad m=0,1,\dots,
\end{eqnarray}
\begin{equation}
\Delta_{-1}\epsilon_p=i(1+q)\sum_{k=0}^\infty (-q)^k b^{(k)}
+i(1+q)\sum_{m=0}^\infty \frac{m!(-1)^{m+1}}{z^{m+1}}
\sum_{k=0}^\infty (-q)^k \tilde R^{(k)}_m.
\label{Deltaepsilonp0}
\end{equation}
We can compare this expression to the assumed product form
\begin{equation}
\Delta_{-1}\epsilon_p=iX\epsilon_p,  
\label{Deltaepsilonp}
\end{equation}
where
\begin{equation}
X=b+\sum_{m=0}^\infty \frac{X_m}{z^{m+1}}.
\end{equation}
Before making the comparison it is useful to recast (\ref{Deltaepsilonp}) in
the form 
\begin{equation}
X\epsilon_p=X+(1+q)\sum_{n=0}^\infty\frac{\phi_{n+1}(q)}{z^{n+1}},
\end{equation}
where 
\begin{equation}
X\sum_{n=0}^\infty \frac{E_{n+1}(q)}{z^{n+1}}
=\sum_{n=0}^\infty \frac{\phi_{n+1}(q)}{z^{n+1}}.
\end{equation}
The comparison of (\ref{Deltaepsilonp0}) with (\ref{Deltaepsilonp}) gives
\begin{equation}
\sum_{k=0}^\infty (-q)^k b^{(k)}=\frac{b}{1+q}=b\sum_{k=0}^\infty (-q)^k,  
\end{equation}
\begin{equation}
\sum_{k=0}^\infty (-q)^k \tilde R^{(k)}_m=\frac{(-1)^{m+1}}{m!}\left\{
\frac{X_m}{1+q}+\phi_{m+1}(q)\right\}.  
\label{comp2}
\end{equation}
The first of these gives
\begin{equation}
b^{(k)}=b,\qquad k=0,1,\dots
\end{equation}
This is what we obtained already from the consistency of asymptotic expansions.
Before using the second comparison (\ref{comp2}) it is useful to write out
$\phi_{m+1}(q)$ explicitly:
\begin{equation}
\phi_{m+1}(q)=bE_{m+1}(q)+\sum_{r=0}^{m-1} X_r E_{m-r}(q)=
b\sum_{k=0}^m p_{m,k}q^k +\sum_{k=0}^{m-1} \sum_{r=0}^{m-1-k} X_r p_{m-r-1,k}q^k.  
\end{equation}
We can now read off from (\ref{comp2}) the relations
(\ref{comp2a})-(\ref{comp2c}).
The structure (\ref{comp2a}) is precisely the same as we already found.
Writing out (\ref{comp2b}) and (\ref{comp2c}) explicitly for $m=0,1,2$ we get
\begin{equation}
\begin{split}
\tilde R^{(0)}_0&=\tilde R_0-p_{0,0}b,\\
\tilde R^{(0)}_1&=\tilde R_1+b p_{1,0}+X_0 p_{0,0}=
\tilde R_1-p_{0,0}\tilde R_0+p_{1,0}b,\\
\tilde R^{(1)}_1&=\tilde R_1-p_{1,1}b,\\
\tilde R^{(0)}_2&=
\tilde R_2-\frac{1}{2}\left\{b p_{2,0}+X_0 p_{1,0}+X_1 p_{0,0}\right\}\\
&=\tilde R_2-\frac{1}{2}p_{0,0}\tilde R_1
+\frac{1}{2}p_{1,0}\tilde R_0
-\frac{1}{2}p_{2,0}b,\\
\tilde R^{(1)}_2&=
\tilde R_2+\frac{1}{2}\left\{b p_{2,1}+X_0 p_{1,1}\right\}=
\tilde R_2-\frac{1}{2}p_{1,1}\tilde R_0+\frac{1}{2}p_{2,1}b,\\
\tilde R^{(2)}_2&=\tilde R_2-\frac{1}{2}p_{2,2}b.
\end{split}
\label{tilderesp2}
\end{equation}
Here we used the identifications (\ref{comp2a}). We see that (\ref{tilderesp2})
exactly reproduces the earlier result (\ref{tilderesp}).

The actual values of the generic coefficients are given by (\ref{irreg}):
\begin{equation}
X_m=2W_m,\qquad m=0,1,\dots
\label{Xm2Wm}  
\end{equation}
Thus the coefficient function is
\begin{equation}
X=1+\sum_{m=0}^\infty \frac{2W_m}{z^{m+1}}=\epsilon  
\end{equation}
and
\begin{equation}
\Delta_{-1}\epsilon_p=i\epsilon\epsilon_p.
\end{equation}

\section{Asymptotic expansion of $\rho$}
\label{appE}

The asymptotic expansion (\ref{ASYYMn}) can also be used to obtain the
asymptotic series for $\rho$.
Using the inversion formula
\begin{equation}
w_{0,M}=\sum_{n=0}^M(-1)^n E_{1,n} 2^{n-M} Y^{(M)}_n  
\end{equation}
we have
\begin{equation}
\tilde u_M=2^Mc_{0,M}=2^M w_{0,M}=\sum_{n=0}^M \omega_n Y^{(M)}_n,\qquad
\omega_n=(-2)^n E_{1,n}.  
\end{equation}
Here are the first few $\omega_n$ coefficients:
\begin{equation}
\omega_0=1,\qquad \omega_1=\frac{3}{4},\qquad \omega_2=-\frac{15}{32},\qquad  
\omega_3=\frac{105}{128}
\end{equation}
and the first few $\tilde u_n$ values:
\begin{equation}
\tilde u_0=1,\qquad \tilde u_1=a-\frac{3}{4},\qquad
\tilde u_2=-\frac{15}{32}+\frac{3a}{4}-\frac{a^2}{2},  
\end{equation}
\begin{equation}
\tilde u_3=-\frac{105}{128}+\frac{45a}{32}-\frac{9a^2}{8}+\frac{a^3}{2}
+\frac{3}{8}\zeta_3.  
\end{equation}
We also recall the first few input parameters:
\begin{equation}
a_0=1,\qquad a_1=a-1,\qquad a_2=1-a+\frac{a^2}{2}-\frac{1}{4}\zeta_2,  
\end{equation}
\begin{equation}
a_3=a-1-\frac{a^2}{2}+\frac{a^3}{6}+\frac{1-a}{4}\zeta_2+\frac{1}{4}\zeta_3.
\end{equation}

The Taylor coefficients of the Borel transform of $\rho$ are
\begin{equation}
c^{(\rho)}_{M-1}=\frac{\tilde u_M}{\Gamma(M)}=\sum_{n=0}^M\omega_n
\frac{Y^{(M)}_n}{\Gamma(M)}.  
\end{equation}

To a given order of the asymptotic expansion (here we again work up to NNNLO)
only a finite number of terms contribute to (\ref{ASYYMn}):
\begin{equation}
\begin{split}
\frac{Y^{(M)}_n}{\Gamma(M)}&=\frac{(-1)^{M-n}}{2\pi(M-1)(\dots)(M-n)}\Big\{
-a_n\Big[b+\frac{R^{(0)}_0}{M-n}\\
&+\frac{R^{(0)}_1}{(M-n)(M-n-1)}
+\frac{2R^{(0)}_2}{(M-n)(M-n-1)(M-n-2)}\Big]\\  
&-a_{n+1}\Big[\frac{R^{(0)}_0-R_0}{M-n}
+\frac{R^{(0)}_1-R^{(1)}_1}{(M-n)(M-n-1)}\\
&+\frac{2[R^{(0)}_2-R^{(1)}_2]}{(M-n)(M-n-1)(M-n-2)}\Big]\\
&+a_{n+2}\Big[\frac{R^{(1)}_1-R_1}{(M-n)(M-n-1)}
+\frac{2[R^{(1)}_2-R^{(2)}_2]}{(M-n)(M-n-1)(M-n-2)}\Big]\\
&-a_{n+3}\Big[\frac{2[R^{(2)}_2-R_2]}{(M-n)(M-n-1)(M-n-2)}\Big]
+\dots\Big\}.  
\label{ASYYMn2}
\end{split}    
\end{equation}
To NNNLO we will need the $n=0,1,2,3$ cases and taking into account the shift
of the index $M$ by 1 unit we have
\begin{equation}
\begin{split}  
\frac{Y^{(M+1)}_0}{\Gamma(M+1)}&=\frac{(-1)^M}{2\pi}\Big\{b+
\frac{(1+a_1)R^{(0)}_0-a_1R_0}{M}\\
&+\frac{1}{M(M-1)}\big[(1+a_1)(R^{(0)}_1-R^{(0)}_0)-(a_1+a_2)R^{(1)}_1+a_1R_0+
a_2 R_1\big]\\
&+\frac{2}{M(M-1)(M-2)}
\big[(1+a_1)(R^{(0)}_2-R^{(0)}_1+R^{(0)}_0)-(a_1+a_2)(R^{(1)}_2-R^{(1)}_1)\\
&+(a_2+a_3)R^{(2)}_2-a_1R_0-a_2R_1-a_3R_2\big]+\dots\Big\},
\end{split}
\end{equation}
\begin{equation}
\begin{split}  
\frac{Y^{(M+1)}_1}{\Gamma(M+1)}&=\frac{(-1)^M}{2\pi}\Big\{
-\frac{ba_1}{M}+\frac{a_2 R_0-(a_1+a_2)R^{(0)}_0}{M(M-1)}\\
&+\frac{(a_1+a_2)(R^{(0)}_0-R^{(0)}_1)+(a_2+a_3)R^{(1)}_1-a_2R_0-a_3R_1}
{M(M-1)(M-2)}+\dots\Big\},
\end{split}
\end{equation}
\begin{equation}
\frac{Y^{(M+1)}_2}{\Gamma(M+1)}=\frac{(-1)^M}{2\pi}\Big\{
\frac{ba_2}{M(M-1)}+\frac{(a_2+a_3)R^{(0)}_0-a_3R_0}{M(M-1)(M-2)}
+\dots\Big\},
\end{equation}
\begin{equation}
\frac{Y^{(M+1)}_3}{\Gamma(M+1)}=\frac{(-1)^M}{2\pi}\Big\{
-\frac{ba_3}{M(M-1)(M-2)}+\dots\Big\}.
\end{equation}

Comparing the expansion of $c^{(\rho)}_n$ to the generic case (\ref{ASYcn})
we find for the constants determining the alien derivative of $\rho$:
\begin{equation}
\begin{split}
\pi\tilde B_o&=b,\\
\pi\tilde B_o\tilde q_0&=(1+a_1)R^{(0)}_0-a_1R_0-ba_1\omega_1,\\
\pi\tilde B_o\tilde q_1&=(1+a_1)(R^{(0)}_1-R^{(0)}_0)-(a_1+a_2)R^{(1)}_1
+a_1R_0+a_2R_1\\
&+\omega_1[a_2R_0-(a_1+a_2)R^{(0)}_0]+ba_2\omega_2,\\
2\pi\tilde B_o\tilde q_2&=2[(1+a_1)(R^{(0)}_2-R^{(0)}_1+R^{(0)}_0)
-(a_1+a_2)(R^{(1)}_2-R^{(1)}_1)\\
&+(a_2+a_3)R^{(2)}_2-a_1R_0-a_2R_1-a_3R_2]\\
&+\omega_1[(a_1+a_2)(R^{(0)}_0-R^{(0)}_1)+(a_2+a_3)R^{(1)}_1-a_2R_0-a_3R_1]\\
&+\omega_2[(a_2+a_3)R^{(0)}_0-a_3R_0]-ba_3\omega_3.
\end{split}    
\end{equation}
Using the solution (\ref{untilderes}) this simplifies to\footnote{Note that
$\tilde R_0=R_0$, $\tilde R_1=R_1-R_0$, $\tilde R_2=R_2-R_1+R_0$.}  
\begin{equation}
\begin{split}
&\pi\tilde B_o\tilde q_0=\tilde R_0+\left(\frac{3}{4}-a\right)b
=-X_0-\tilde u_1 b,\\
&\pi\tilde B_o\tilde q_1=\tilde R_1+\left(\frac{3}{4}-a\right)\tilde R_0+
\left(-\frac{15}{32}+\frac{3a}{4}-\frac{a^2}{2}\right)b
=X_1+\tilde u_1 X_0+\tilde u_2 b,\\
2&\pi\tilde B_o\tilde q_2=2\tilde R_2+\left(\frac{3}{4}-a\right)\tilde R_1+
\left(-\frac{15}{32}+\frac{3a}{4}-\frac{a^2}{2}\right)\tilde R_0\\
&+\left(\frac{105}{128}-\frac{45a}{32}+\frac{9a^2}{8}-\frac{a^3}{2}
-\frac{3}{8}\zeta_3\right)b
=-X_2-\tilde u_1 X_1-\tilde u_2 X_0-\tilde u_3 b.
\end{split}    
\end{equation}
From here we can read off the alien derivative of $\rho$, given by
(\ref{alienrho}).

\section{Alien derivatives}
\label{appF}

In this appendix we summarize our conventions for alien derivatives.
For details, see \cite{Marino:2012zq,Dorigoni:2014hea,Aniceto:2018bis}.

Let us recall that we start from a formal asymptotic series
\begin{equation}
\Psi(z)=1+\sum_{n=1}^\infty \frac{\Gamma(n) c_{n-1}}{z^n}
\label{Borelcn}
\end{equation}
and define its Borel transform as
\begin{equation}
\hat\Psi(t)=\sum_{n=0}^\infty c_n t^n.  
\end{equation}
Here we assume that the coefficients have the asymptotic expansion
\begin{equation}
\begin{split}  
c_n=&\frac{\tilde A_o}{2}\Big\{1+\frac{\tilde p_0}{n}+\frac{\tilde p_1}{n(n-1)}
+\frac{2\tilde p_2}{n(n-1)(n-2)}+\dots\Big\}\\
+(-1)^n&\frac{\tilde B_o}{2}\Big\{1+\frac{\tilde q_0}{n}
+\frac{\tilde q_1}{n(n-1)}+\frac{2\tilde q_2}{n(n-1)(n-2)}+\dots\Big\}.
\end{split}
\label{ASYcn}
\end{equation}
In this case the singular part of the Borel transform is
\begin{equation}
\begin{split}  
\hat\Psi^{\rm sing}(t)&=\frac{\tilde A_o}{2}\Big\{\frac{1}{1-t}
-\ln(1-t)\sum_{m=0}^\infty \tilde p_m(t-1)^m\Big\}\\
&+\frac{\tilde B_o}{2}\Big\{\frac{1}{1+t}
-\ln(1+t)\sum_{m=0}^\infty (-1)^m \tilde q_m(t+1)^m\Big\}
\end{split}
\end{equation}
and the alien derivatives are given by
\begin{equation}
\begin{split}
\Delta_1\Psi(z)&=-i\pi\tilde A_o\Big\{1+\sum_{m=0}^\infty \frac{m!\tilde p_m}
{z^{m+1}}\Big\},\\
\Delta_{-1}\Psi(z)&=i\pi\tilde B_o\Big\{1-\sum_{m=0}^\infty
\frac{m!(-1)^m\tilde q_m}{z^{m+1}}\Big\}.
\label{alien}
\end{split}
\end{equation}

\section{The $\eta_k=1$ model}
\label{eta1}

Since all $\eta_k$ are algebraically independent, it makes sense to study the
simplest model
\begin{equation}
\eta_k=1,\qquad k=1,2,\dots
\end{equation}
For this model
\begin{equation}
g(x)=1,\qquad a_n=\delta_{n,0}.  
\end{equation}

Let us make the following definitions.
\begin{equation}
K_n=p_{n-1,0},\quad n=1,2,\dots,\qquad\quad K_0=1  
\end{equation}
and
\begin{equation}
S_n=2^nE_{0,n}=\frac{1}{n!}\prod_{\ell=1}^n \left(\ell-\frac{1}{2}\right)^2,\quad
n=1,2\dots,\qquad\quad S_0=1.  
\end{equation}
In this simple model the building blocks simplify enormously:
\begin{equation}
Y^{(M)}_n=\delta_{n,0} K_M,\quad n=0,\dots,M,  
\end{equation}
\begin{equation}
w_{a,M}=\delta_{a,0} 2^{-M} K_M,\quad M=0,1,\dots,\quad a=0,\dots,M.
\end{equation}
The left and right sides of the Volin equations simplify accordingly to
\begin{equation}
{\cal R}^{(1)}_1=p_{0,0}
\end{equation}
and
\begin{equation}
M\geq2\quad\left\{  
\begin{split}
{\cal R}^{(M)}_r&=p_{M-1,r-1}+p_{M-1,r},\quad r=1,\dots,M-1,\\
{\cal R}^{(M)}_M&=p_{M-1,M-1},
\end{split}
\right.
\end{equation}
\begin{equation}
{\cal L}^{(M)}_r=S_rK_{M-r}.
\end{equation}
Equating the two sides we obtain
\begin{eqnarray}
&&M\geq1\qquad p_{M-1,M-1}=S_M,\\
&&M\geq2\qquad p_{M-1,r-1}+p_{M-1,r}=S_r K_{M-r},\quad r=1,\dots,M-1.
\label{precur2}
\end{eqnarray}
It is easy to find the solution of this recursion:
\begin{equation}
p_{M-1,M-k}=\sum_{p=0}^{k-1}(-1)^{k-p+1} K_p S_{M-p},\qquad k=1,\dots,M.  
\end{equation}
In particular, the most interesting unknowns $K_n$ are determined recursively as
\begin{equation}
K_{M+1}=\sum_{p=0}^M (-1)^{M-p} K_p S_{M+1-p},\qquad M=0,1,\dots  
\label{Krecur}
\end{equation}
Summing up (\ref{precur2}) from $r=1$ to $r=M-1$ we obtain 
\begin{equation}
2W_{M-1}=\sum_{r=0}^M S_r K_{M-r},  
\end{equation}
which also holds for $M=1$. If we define
\begin{equation}
\xi_M=\sum_{r=0}^M S_r K_{M-r},\quad
\label{xidef}
\end{equation}
we can write
\begin{equation}
\epsilon(x)=\sum_{n=0}^\infty \xi_n x^n.  
\end{equation}
Here $x=1/z$. Similarly
\begin{equation}
\rho(x)=\sum_{n=0}^\infty K_n x^n  
\end{equation}
and we also introduce
\begin{equation}
B(x)=\sum_{n=0}^\infty S_n x^n.  
\end{equation}
The solution of the recursion (\ref{Krecur}) is summarized by
\begin{equation}
\rho(x)B(-x)=1  
\end{equation}
and (\ref{xidef}) is equivalent to
\begin{equation}
\epsilon(x)=B(x)\rho(x).  
\end{equation}
Let us introduce
\begin{equation}
B(x)=1+b(x),\qquad b(x)=\sum_{n=1}^\infty S_n x^n.  
\end{equation}
The Borel transform of $b(x)$ is
\begin{equation}
\hat b(t)=\sum_{n=0}^\infty \beta_n t^n,  
\end{equation}
where
\begin{equation}
\beta_n=\frac{S_{n+1}}{\Gamma(n+1)},\quad n=0,1,\dots
\label{betan}
\end{equation}
We also define
\begin{equation}
D(x)=B(-x)=1+d(x),\qquad d(x)=b(-x)  
\end{equation}
and see that
\begin{equation}
\hat d(t)=-\hat b(-t).
\end{equation}
By inspecting the Taylor expansion coefficients (\ref{betan}) we recognize that
\begin{equation}
\hat b(t)=\frac{1}{4}{}_2F_1\left(\frac{3}{2},\frac{3}{2};2;t\right).  
\end{equation}
Using {\tt Mathematica} we can verify that in the vicinity of $t=1$
\begin{equation}
\hat b(t)=-\frac{1}{\pi(t-1)}-\frac{\ln(1-t)}{\pi} \hat d(t-1)+\Phi(t-1),  
\label{singb}
\end{equation}
where $\Phi$ is regular around the origin. This can be written in the language
of alien derivatives as
\begin{equation}
\Delta_1 B=-2i-2id=-2iD.  
\end{equation}
Similarly we have
\begin{equation}
\Delta_{-1}D=-2i B.  
\end{equation}
From
\begin{equation}
\rho=\frac{1}{D},\qquad\quad \epsilon=\frac{B}{D}
\end{equation}
we calculate
\begin{equation}
\Delta_1\rho=0,\qquad\quad \Delta_1\epsilon=\frac{-2iD}{D}=-2i,  
\end{equation}
\begin{equation}
\Delta_{-1}\rho=-\frac{1}{D^2}\Delta_{-1}D=\frac{2iB}{D^2}=2i\epsilon\rho,
\end{equation}
\begin{equation}
\Delta_{-1}\epsilon=-\frac{B}{D^2}\Delta_{-1}D=\frac{2iB^2}{D^2}=2i\epsilon^2.
\end{equation}
This structure of alien derivatives is very similar to what was found
in the main part of the paper for the full model, it is only that $\Delta_1$
is by a factor 2 too small
and $\Delta_{-1}$ is by a factor 2 too large. However, the overall factors can
be changed easily, as we will see in the next subsection.

\subsection{$a$-dependence}

The alien derivatives transform under a change
of variables. If the original expansion parameter $z$ is changed to $x$, where
\begin{equation}
z=z(x),  
\end{equation}
then the asymptotic expansions $\gamma(z)$ are changed to $C(x)$ by the
simple substitution rule
\begin{equation}
C(x)=\gamma(z(x)).  
\end{equation}
Let us denote the original alien derivative (obtained from the Borel transform
of $\gamma(z)$) at $\omega$ by $\Delta_\omega\gamma$ and the alien derivative
in the new variable (obtained from the Borel transform of $C(x)$) by
$D_\omega C$. The transformation rule $\Delta_\omega\longrightarrow D_\omega$
is given by \cite{composit}
\begin{equation}
D_{\omega}\gamma(z(x))={\rm e}^{-\omega(z(x)-x)}
(\Delta_{\omega}\gamma)(z(x))+\gamma^{\prime}(z(x))(D_{\omega}z)(x).
\label{saeedeh}
\end{equation}
In the case of a simple shift by a constant $m$,
\begin{equation}
z(x)=x+m,  
\end{equation}
(\ref{saeedeh}) reduces to a constant rescaling:
\begin{equation}
D_\omega C={\rm e}^{-m\omega}\Delta_\omega \gamma.
\end{equation}
If we treat the symbol $a$ in (\ref{gx}) as a generic parameter (rather then
a shorthand for its true numerical value $a=\ln2$) then
\begin{equation}
\epsilon(a,z)\qquad\quad {\rm and}\qquad\quad s(a,z)=z\rho^2(a,z)
\end{equation}
have the following remarkable property\footnote{These hold for
arbitrary $\eta_k$, not just for the simplified model
$\eta_k=1$. We have verified this property perturbatively up to $16^{\rm th}$
order using Mathematica.}:
\begin{equation}
s(a,z-2a)=s_o(z),\qquad\qquad \epsilon(a,z-2a)=\epsilon_o(z),  
\end{equation}
where $s_o(z)$ and $\epsilon_o(z)$ are independent of the parameter $a$.

In the simplified model we calculated
$\epsilon(1,z)$ and $\rho(1,z)$. For example for $\epsilon$ we then have  
\begin{equation}
\epsilon(a,x)=\epsilon(1,x+2a-2)
\end{equation}
and
\begin{equation}
D_\omega\epsilon={\rm e}^{(2-2a)\omega}\Delta_\omega\epsilon=
\left(\frac{{\rm e}}{2}\right)^{2\omega}\Delta_\omega\epsilon.  
\end{equation}
The transformation for the alien derivatives of $\rho$ is the same.

We see that after restoring $a=\ln2$ the alien derivatives of $\epsilon$
change to
\begin{equation}
D_1\epsilon=-i({\rm e}^2/2),\qquad\qquad
D_{-1}\epsilon=i(8/{\rm e}^2)\epsilon^2.
\end{equation}
The changes for the alien derivatives of $\rho$ are the same. It is remarkable
that these coefficients are numerically quite close to the \lq\lq true''
values (4 and 1):
\begin{equation}
D_1\epsilon=-3.69i,\qquad\qquad
D_{-1}\epsilon=1.08 i \epsilon^2.
\end{equation}

\subsection{The alien derivatives of $\epsilon_p$ for the $\eta_k=1$ model}

We calculate $\epsilon_p$ for $\eta_k=1$ as follows. Let us multiply
(\ref{precur2}) by $q^r$ and sum over $r$ from $1$ to $M-1$. In this way
we obtain the relation (for $M\geq2$) 
\begin{equation}
\sum_{r=0}^M S_r K_{M-r} q^r=(1+q) E_M(q).  
\end{equation}
The above relation also holds for $M=1$ and we can write
\begin{equation}
\begin{split}
\epsilon_p&=1+(1+q)\sum_{M=1}^\infty \frac{E_M(q)}{z^M}=\sum_{M=0}^\infty
\frac{1}{z^M}\sum_{r=0}^M S_r K_{M-r} q^r\\
&=\sum_{r=0}^\infty \sum_{k=0}^\infty x^{r+k} S_r K_k q^r=\rho(x)B(qx).
\end{split}     
\end{equation}
For clarity let us define
\begin{equation}
P(x)=B(qx)=1+p(x)=1+\sum_{n=1}^\infty S_n q^n x^n.  
\end{equation}
The Borel transform of $p(x)$ is
\begin{equation}
\hat p(t)=q\hat b(qt)  
\end{equation}
and therefore its singular part, using (\ref{singb}), is
\begin{equation}
\begin{split}  
\hat p(t)&\approx \frac{-q}{\pi(qt-1)}-\frac{q\ln(1-qt)}{\pi}\hat d(qt-1)
+\dots\\   
&\approx \frac{-1}{\pi(t-p)}-\frac{\ln(p-t)}{\pi}q\hat d(q(t-p))+\dots
\end{split}
\end{equation}
From here we can read off the alien derivative
\begin{equation}
\Delta_p P=-2i-2id(qx)=-2iD(qx).  
\end{equation}
Now we can calculate
\begin{equation}
\Delta_p\epsilon_p=\rho(x)\Delta_p P(x)=-2i\rho(x)D(qx)=-2i\epsilon_{-p}  
\end{equation}
and
\begin{equation}
\Delta_{-1}\epsilon_p=B(qx)\Delta_{-1}\rho=2i\epsilon\rho
B(qx)=2i\epsilon\epsilon_p.
\end{equation}

\bibliographystyle{JHEP}
\bibliography{borel}

\end{document}